\documentclass[12pt]{book}
\input{epsf}
\usepackage{cite}
\usepackage{amsmath}
\usepackage{graphicx}
\usepackage{bm}
\def\etal{{\it et al.\ }}
\def\ltwid{\mathrel{\raise.3ex\hbox{$<$\kern-.75em\lower1ex\hbox{$\sim$}}}}

\begin{document}

\setlength{\baselineskip}{.3in}

\chapter*{Numerical Studies of the 2D Hubbard Model}
\centerline{\large\bf D.J.~Scalapino}
\centerline{Department of Physics, University of California, Santa Barbara, CA 93106-9530, USA}

\section*{Abstract}

Numerical studies of the two-dimensional Hubbard model have shown that it exhibits
the basic phenomena seen in the cuprate materials.  At half-filling one finds an
antiferromagnetic Mott-Hubbard groundstate.  When it is doped, a pseudogap appears
and at low temperature d-wave pairing and striped states are seen.  In addition, there
is a delicate balance between these various phases.  Here we review evidence for this and
then discuss what numerical studies tell us about the structure of the interaction which
is responsible for pairing in this model.

\section*{1. Introduction}

A variety of numerical methods have been used to study Hubbard and t-J models and there
are a number of excellent reviews.\cite{ref:1,ref:2,ref:3,ref:4,ref:5,ref:6,ref:7,ref:DR}
The approaches have ranged from Lanczos diagonalization\cite{ref:1,ref:2,ref:8,ref:A1,ref:9}
of small clusters to
density-matrix-renormalization-group studies of n-leg ladders\cite{ref:DR,ref:10,ref:B1,ref:29}
and quantum Monte Carlo simulations of two-dimensional
lattices\cite{ref:3,ref:11,ref:12,ref:A2,ref:A3,ref:Scal1,ref:A4,ref:A5,ref:A6,ref:13,ref:A6a}.
In addition, recent cluster generalizations of dynamic mean-field
theory\cite{ref:4,ref:6,ref:7,ref:MP,ref:14,ref:16,ref:37,ref:15,ref:17,ref:SSK,ref:B2,ref:LK}
are providing new insight into the low temperature properties of these models. A
significant finding of these numerical studies is that these basic models can exhibit
antiferromagnetism, stripes, pseudogap behavior, and $d_{x^2-y^2}$ pairing.  In addition,
the numerical studies have shown how delicately balanced these models are
between nearly degenerate phases.  Doping away from half-filling can tip the
balance from antiferromagnetism to a striped state in which half-filled domain walls
separate $\pi$-phase-shifted antiferromagnetic regions.  Altering the next-near-neighbor
hopping $t^\prime$ or the strength of $U$ can favor $d_{x^2-y^2}$ pairing correlations over
stripes.  This delicate balance is also seen in the different results obtained using
different numerical techniques for the same model. For example, density matrix renormalization
group (DMRG) calculations for doped 8-leg t-J ladders find evidence for a striped ground
state.\cite{ref:10} However, variational and Green's function Monte Carlo calculations for the
doped t-J lattice, pioneered by Sorella and co-workers,\cite{ref:13,ref:A6a} find
groundstates characterized by $d_{x^2-y^2}$ superconducting order with only weak signs
of stripes.  Similarly, DMRG calculations for doped 6-leg Hubbard ladders \cite{ref:29}
find stripes when the ratio of $U$ to the near-neighbor hopping $t$ is greater than 3,
while various cluster calculations \cite{ref:16,ref:17,ref:SSK,ref:B2,ref:LK} find evidence that
antiferromagnetism and $d_{x^2-y^2}$ superconductivity compete in this same parameter regime.
These techniques represent present day state-of-the-art numerical approaches.  The fact
that they can give different results may reflect the influence of different boundary
conditions or different aspect ratios of the lattices that were studied.  The $n$-leg
open boundary conditions in the DMRG calculations can favor stripe formation. Alternately,
the cluster lattice sizes and boundary conditions can frustrate stripe formation. It is also
possible that these differences reflect subtle numerical biases in the different numerical
methods. Nevertheless, these results taken together show that both the striped and the
$d_{x^2-y^2}$ superconducting phases are nearly degenerate low energy states of the doped
system.  Determinantal quantum Monte Carlo calculations\cite{ref:A5} as well as various
cluster calculations show that the underdoped Hubbard model also exhibits pseudogap
phenomena.\cite{ref:B2,ref:37,ref:15,ref:16,ref:SSK,ref:17} The remarkable similarity of this
behavior to the range of phenomena observed in the cuprates provides strong evidence
that the Hubbard and t-J models indeed contain a significant amount of the essential
physics of the problem.\cite{ref:19}

In this chapter, we will focus on the one-band Hubbard model.  Section~2 provides an overview
of the numerical methods which were used to obtain the results that will be discussed.
We have selected three methods, determinantal quantum Monte Carlo, the dynamic cluster
approximation and the density-matrix-renormalization group.  In principle, these methods
provide unbiased approaches which can be extrapolated to the bulk limit or in
the case of the DMRG, to ``infinite length" ladders. This choice has left out many other
important techniques such as the zero variance extrapolation of projector Monte
Carlo,\cite{ref:13,ref:A6a}, variational cluster
approximations,\cite{ref:MP,ref:14,ref:15,ref:17,ref:SSK,ref:B2} renormalization group
flow techniques,\cite{ref:MS,ref:RNG1,ref:34} high temperature series expansions
\cite{ref:PL,ref:PKZ,ref:KO} and the list undoubtedly goes on.  It was motivated by
the need to write about methods with which I have direct experience.

In Section~3 we review the numerical evidence showing that the Hubbard model can indeed
exhibit antiferromagnetic, $d_{x^2-y^2}$ pairing and striped low-lying states as well as
pseudogap phenomena.  From this we conclude that the Hubbard model provides a basic
description of the cuprates, so that the next question is what is the interaction that
leads to pairing in this model?  From a numerical approach, this question is different from
determining whether the groundstate is antiferromagnetic, striped, or superconducting.
Here, one would like to understand the structure of the underlying effective
interaction that leads to pairing.  Although on the surface this might seem like a more
difficult question to address numerically, it is in fact easier than determining the
nature of the long-range order of the low temperature phase.  The phase determination
problem involves an extrapolation to an infinite lattice at low, or in the case of
antiferromagnetism, to zero temperature. However, the pairing
interaction is short-ranged and is formed at a temperature above the
superconducting transition so that it can be studied on smaller clusters and
at higher temperatures.

As discussed in Section~4, the effective pairing interaction is given by the
irreducible particle-particle vertex $\Gamma^{\rm pp}$.  Using Monte Carlo results for
the one- and two-particle Green's functions, one can determine $\Gamma^{\rm pp}$ and
examine its momentum and Matsubara frequency dependence.\cite{ref:20,ref:21}  One can also determine if it
is primarily mediated by a particle-hole magnetic ($S=1$) exchange, a charge density
($S=0$) exchange, or by a more complex mechanism. Section 5 contains a summary and
our conclusions.

\section*{2. Numerical Techniques}

In this chapter we will be reviewing numerical results which have been obtained for the 2D
Hubbard model. It would, of course, be simplest if one could say that these results were
obtained by ``exact'' diagonalization on a sequence of L$\times$L lattices with a
``finite-size scaling'' analysis used to determine the bulk limit. While one might not know
the exact details of how this was done, one understands what it means.  Unfortunately the
exponential growth of the Hilbert space with lattice size limits this approach and other
less familiar and often less transparent methods are required.

In this chapter, we will discuss results obtained using the determinantal quantum Monte
Carlo algorithm,\cite{ref:22,ref:12}
a dynamic cluster approximation,\cite{ref:6,ref:16} and the density matrix renormalization
group.\cite[5]{ref:24,ref:25} All of these
methods have been described in detail in the literature. However, to provide a context for
the numerical results discussed in Sections 3 and 4, we proceed with a brief overview of these
techniques.

\subsection*{Determinantal Quantum Monte Carlo}

The determinantal quantum Monte Carlo method was introduced in order to numerically
calculate finite temperature expectation values.
\begin{equation}
\langle A\rangle = \frac{Tr\, e^{-\beta H} A}{z}\, .
\label{one}
\end{equation}
Here, $H$ includes $-\mu N$ so that $Z=Tr\, e^{-\beta H}$ is the grand partition function.
For the Hubbard model, the Hamiltonian is separated into a one-body term
\begin{equation}
K=-t \sum_{\langle ij\rangle \sigma} \left(c^\dagger_{i\sigma} c_{j\sigma} +
c^\dagger_{j\sigma} c_{i\sigma}\right) - \mu \sum_{i\sigma} n_{i\sigma}
\label{two}
\end{equation}
and an interaction term
\begin{equation}
V=U \sum_i \left(n_{i\uparrow}-\frac{1}{2}\right)\, \left(n_{i\downarrow}-\frac{1}{2}\right)
\, .
\label{three}
\end{equation}
Then, dividing the imaginary time interval $(0,\beta)$ into M segments of width
$\Delta\tau$, we have
\begin{equation}
e^{-\beta H} = \left[e^{-\Delta\tau(K+V)}\right]^M \simeq \left[e^{-\Delta\tau K}
e^{-\Delta\tau V}\right]^M\, .
\label{four}
\end{equation}
In the last term, a Trotter breakup has been used to separate the non-commuting operators
$K$ and $V$. This leads to errors of order $tU\Delta \tau^2$ which can be systematically
treated by reducing the size of the time slice $\Delta\tau$. Then, on each
$\tau_\ell=\ell\Delta\tau$ slice and for each lattice site $i$, a discrete
Hubbard-Stratonovich field\cite{ref:26} $S_i(\tau_\ell)=\pm 1$ is introduced so that the interaction can
be written as a one-body term
\begin{equation}
e^{-\Delta\tau U\left(n_{i\uparrow}-\frac{1}{2}\right)\,
\left(n_{i\downarrow}-\frac{1}{2}\right)}= \frac{e^{-\frac{\Delta\tau U}{4}}}{2}
\sum_{S_i(\tau_\ell)=\pm 1} e^{-\Delta \tau\lambda S_i(\tau_\ell)\,
\left(n_{i\uparrow}-n_{i\downarrow}\right)}
\label{five}
\end{equation}
with $\lambda$ set by $\cosh (\Delta\tau\lambda)=\exp (\Delta\tau \frac{U}{2})$.  The
interacting electron problem has now been replaced by the problem of many electrons
coupled to a $\tau$-dependent Hubbard-Stratonovich Ising field $S_i(\tau_\ell)$ which is to
be averaged over. This average will be done by Monte Carlo importance sampling.

For an $L\times L$ lattice, it is useful to introduce a one-body $L^2\times L^2$ lattice
Hamiltonian $h_\sigma(S(\tau_\ell))$ for spin
$\sigma$ electrons interacting with the Hubbard-Stratonovich field on the
$\tau_\ell$-imaginary time slice
\begin{eqnarray}
\sum_{\langle ij\rangle} c^\dagger_{i\sigma} h_\sigma (S(\tau_\ell))c_{j\sigma} & = & - t
\sum\limits_{\langle ij\rangle} \left(c^\dagger_{i\sigma} c_{j\sigma} + c^\dagger_{j\sigma}
c_{i\sigma}\right)\nonumber\\
 - \mu \sum\limits_i n_{i\sigma} & \pm & \lambda \sum\limits_i S_i(\tau_\ell)
\, n_{i\sigma}\, .
\label{six}
\end{eqnarray}
The plus sign is for spin-up $(\sigma=1)$ and the minus sign is for spin-down
$(\sigma=-1)$. Then, tracing out the fermion degrees of freedom, one obtains
\begin{equation}
Z=\sum_{\{S\}} {\rm det}\, M_\uparrow (S)\, {\rm det}\, M_\downarrow (S)\, .
\label{seven}
\end{equation}
The sum is over all configurations of the $S_i(\tau_\ell)$ field and $M_\sigma (S)$ is an
$L^2\times L^2$ matrix which depends upon this field,
\begin{equation}
M_\sigma (S) = I+ B^\sigma_M B^\sigma_{M-1} \cdots B_1^\sigma\, .
\label{eight}
\end{equation}
$I$ is the unit $L^2\times L^2$ matrix and $B^\sigma_\ell=e^{-\Delta\tau
h_\sigma(S(\tau_\ell))}$ acts as an imaginary time propagator which evolves a state
from $(\ell-1)\, \Delta\tau$ to $\ell\Delta\tau$.

The expectation value $\langle A\rangle$ becomes
\begin{equation}
\langle A\rangle = \sum_{\{S\}} A(S)\ \frac{{\rm det}\, M_\uparrow (S)\, {\rm det}\,
M_\downarrow (S)}{Z}
\label{nine}
\end{equation}
with $A(S)$ the estimator for the operation  $A$ which depends only upon the
Hubbard-Stratonovich field. Typically, we are interested in Green's functions.
For example, the estimator for the one-electron Green's function is
\begin{equation}
G_{ij\sigma} (\tau_\ell, (S)) = \frac{1}{1+B^\sigma_M B^\sigma_{M-1} \cdots B^\sigma_1}
\ B^\sigma_\ell B^\sigma_{\ell-1}\cdots B^\sigma_1
\label{ten}
\end{equation}
and
\begin{equation}
G_{ij\sigma}(\tau_\ell)= \sum_{\{S\}} G_{ij} (T_\ell, (S))\ \frac{{\rm det}\, M_\uparrow
(S)\, {\rm det}\, M_\downarrow(S)}{Z}
\label{eleven}
\end{equation}
The calculations of various susceptibilities and multiparticle Green's functions are
straightforward since once the Hubbard-Stratonovich transformation is introduced, one has a
Wick theorem for the fermion operators.  The only thing that one needs to remember is that
disconnected diagrams must be retained because they can become connected by the subsequent
average over the $S_i(\tau_\ell)$ field.

In computing the product of the $B$ matrices, one must be careful to control round-off
errors as the number of products becomes large at low temperatures or at large $U$ where
$\Delta\tau$ must be small.  In addition, there can be problems inverting the
ill-conditioned sum of the unit matrix $I$ and the product of the $B$ matrices needed in
Eqs.~\eqref{eight} and \eqref{ten}. Fortunately, a matrix stabilization procedure\cite{ref:12} overcomes
these difficulties.

For the half-filled case $(\mu=0)$, provided there is only a near-neighbor hopping, $H$
is invariant under the particle-hole transformation $c_{i\downarrow}=(-1)^i
c^\dagger_{i\downarrow}$. Under this transformation, the last term in Eq.~\eqref{six} for $\sigma=-1$ becomes
\begin{equation}
-\lambda \sum_i S_i(\tau_\ell)\, (1-n_{i\downarrow})
\label{twelve}
\end{equation}
so that
\begin{equation}
{\rm det}\, M_\downarrow (S)=\prod_{i\ell} e^{\lambda\Delta\tau S_i(\tau_\ell)}
{\rm det}\, M_\uparrow (S)
\label{thirteen}
\end{equation}
This means that ${\rm det}\, M_\uparrow (S)\, {\rm det}\, M_\downarrow (S)$ is positive
definite. In this case, the sum over the Hubbard-Stratonovich configurations can be done by
Monte Carlo importance sampling, with the probability of a particular configuration
$\{S(\tau_\ell)\}$ given by
\begin{equation}
P(S)=\frac{{\rm det}\, M_\uparrow (S)\, {\rm det}\, M_\downarrow (S)}{Z}\, .
\label{fourteen}
\end{equation}
Given $M$-independent configurations, selected according to the probability distribution
Eq.~\ref{fourteen}, the expectation value of $A$ is
\begin{equation}
\langle A\rangle = \frac{1}{M} \sum_{\{S\}} A(S)\, .
\label{fifteen}
\end{equation}

When the system is doped away from half-filling, the product ${\rm det}M_\uparrow(S)
\, {\rm det}M_\downarrow (S)$ can become negative.  This is the so-called
``fermion sign problem''.  In this case, one must use the absolute value of
the product of determinants to have a positive definite probability distribution for the
Hubbard-Stratonovich configurations.
\begin{equation}
P_\| (S) = \frac{|{\rm det}\, M_\uparrow (S)\, {\rm det}\, M_\downarrow
(S)|}{\sum\limits_{\{S\}} | {\rm det}\, M_\uparrow (S)\, {\rm det}\, M_\downarrow (S)|}
\label{sixteen}
\end{equation}
Then, in order to obtain the correct results for physical observables, one must include
the sign of the product of determinants\cite{ref:27}
\begin{figure}[htb]
\begin{center}
\includegraphics[height=10cm]{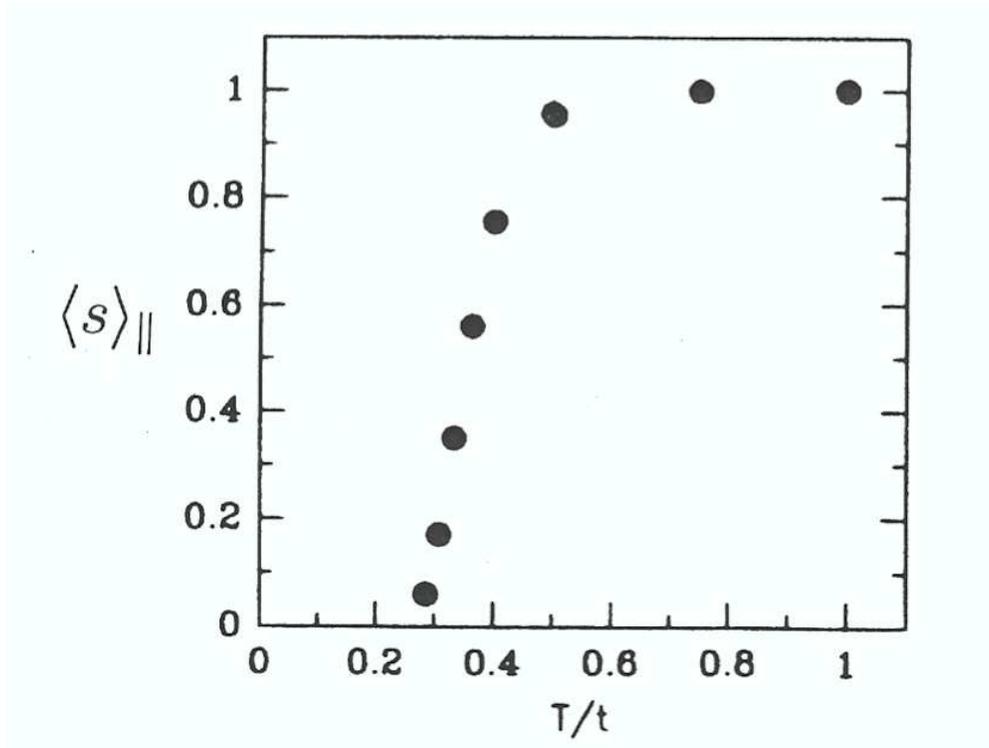}
\caption{The average of the sign of the product of the fermion determinants,
Eq.~\ref{seventeen}, that enters in the determinantal Monte Carlo calculations is shown
versus temperature for an $8\times8$ lattice with $U=8t$ and $\langle n\rangle=0.87$.
One can understand why calculations for $U=8t$ with $T<0.3$ become extremely difficult.
(Scalapino \protect\cite{ref:Scal1})}
\label{fig:1}
\end{center}
\end{figure}
\begin{equation}
s=Sgn ({\rm det}\, M_\uparrow (S)\, {\rm det}\, M_\downarrow (S))
\label{seventeen}
\end{equation}
in the measurement
\begin{equation}
\langle A\rangle = \frac{\langle As\rangle_\|}{\langle s\rangle_\|}\, .
\label{eighteen}
\end{equation}
The $\parallel$
subscript denotes that the average is over configurations generated with the probability
distribution $P_\|$ given by Eq.~\eqref{sixteen}. If the average sign
$\langle s\rangle_\|$ becomes small, there will be large
statistical fluctuations in the Monte Carlo results.  For example,
if $\langle s\rangle_\|=0.1$, one would
have to sample of order $10^2$ times as many independent configurations in order to obtain
the same statistical error as when $\langle s\rangle_\|=1$. On general grounds, one expects
that $\langle s\rangle_\|$
decreases exponentially as the temperature is lowered.

The average sign $\langle s\rangle_\|$ also decreases as $U$ increases and makes it
(exponentially) difficult to obtain results at low temperatures for $U$ of order the
bandwidth and dopings of interest.  Figure~\ref{fig:1} illustrates just how serious
this problem is and why other methods are required.
%initially as the system is doped away from half-filling
%as shown in Fig.~\ref{fig:1}.  Here there are 16 sites and $U=4t$.  The upper part of the
%figure shows how $\langle s\rangle_\|$ varies with the deviation from half-filling
%$\delta=1-\langle n\rangle$ for different values of the temperature.  Results for the
%determinantal quantum Monte Carlo algorithm are labeled QMC.  Results for the dynamic
%cluster approximation are labeled DCA and will be discussed next.  The results
%for $\langle s\rangle_\|$ shown in the lower half of Fig.~\ref{fig:1} are for a 16-site
%lattice with $U=4t$ and $\delta=0.2$  It is unfortunate that the sign problem is the most
%severe in just the doping region of interest.

\subsection*{The Dynamic Cluster Approximation}

In the determinantal quantum Monte Carlo approach, one could imagine carrying out
calculations on a set of L$\times$L lattices and then scaling to the bulk thermodynamic
limit.  The dynamic cluster approximation\cite{ref:6} takes a different approach in which the bulk
lattice is replaced by an effective cluster problem embedded in an external bath designed
to represent the remaining degrees of freedom.  In contrast to numerical studies of
finite-sized systems in which the exact state of an $L\times L$ lattice is determined and
then regarded as an approximation to the bulk thermodynamic result, the cluster theories
give approximate results for the bulk thermodynamic limit.  Then, as the number of cluster sites
increases, the bulk thermodynamic result is approached.

\begin{figure}[htb]
\begin{center}
\includegraphics[width=10cm,angle=0]{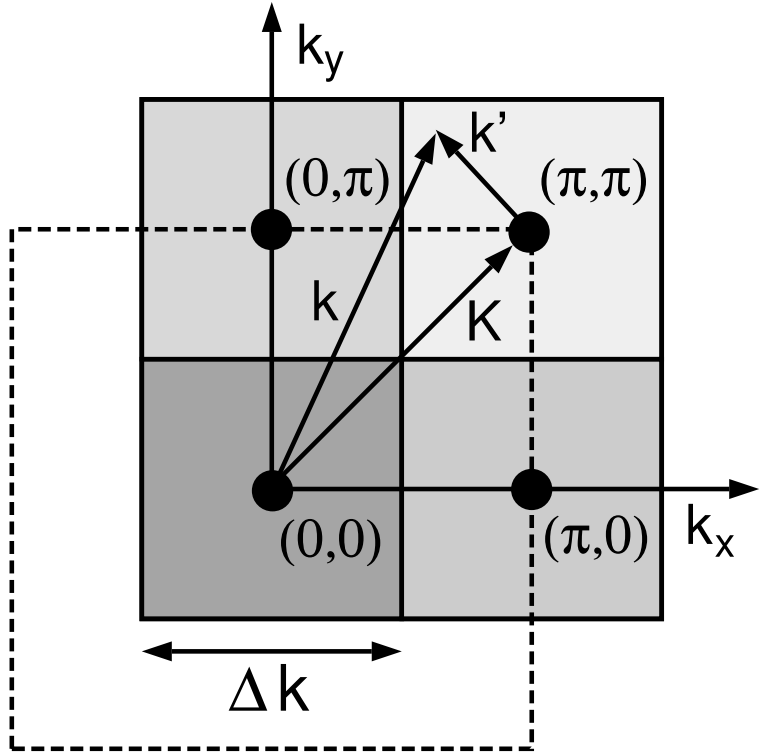}
\caption{In the dynamic cluster approximation the Brillouin zone is divided into $N_c$ cells each represented by a cluster momentum $K$.  Then the self-energy and 4-point vertices are calculated on the cluster using an action determined by the inverse of the coarse-grained cluster-excluded propagator $\mathcal{G}^{-1}$, Eq.~\protect\ref{twenty-one}. This figure illustrates this coarse graining of the Brillouin zone for $N_c=4$ (Maier \etal\protect\cite{ref:6}).}
\label{fig:2}
\end{center}
\end{figure}

In the dynamic cluster approximation, the Brillouin zone is divided into $N_c=L^2$ cells
of size $(2\pi/L)^2$.  As illustrated in Fig.~\ref{fig:2}, each cell is represented by
a cluster momentum ${\bf K}$ placed at the center
of the cell. Then the self-energy $\Sigma({\bf k},\omega_n)$ is approximated by a
coarse grained self-energy
\begin{equation}
\Sigma ({\bf K}+{\bf k}^\prime, w_n) \simeq \Sigma_c({\bf K}, \omega_n)
\label{nineteen}
\end{equation}
for each ${\bf k}^\prime$ within the ${\bf K}^{\rm th}$ cell. The coarse grained Green's function is
given by
\begin{equation}
\bar G({\bf K}, \omega_n)\cong \frac{N_c}{N}\ \sum_{{\bf k}^\prime}
\ \frac{1}{i\, \omega_n-(\varepsilon_{{\bf K}+{\bf k}^\prime}-\mu)-\Sigma_c({\bf K}, \omega)}
\label{twenty}
\end{equation}
where the lattice self-energy is replaced by the coarse grained self-energy. Given
$\bar G$ and $\Sigma_c$, one can set up a quantum Monte Carlo algorithm \cite{ref:6,ref:23} to
calculate the cluster Green's function. Here, the bulk lattice properties are encoded by
using the cluster-excluded inverse Green's function
\begin{equation}
\mathcal{G}^{-1}({\bf K}, \omega_n)=\bar G^{-1} ({\bf K}, \omega_n) + \Sigma_c ({\bf K}, \omega_n)
\label{twenty-one}
\end{equation}
to set up the bilinear part of the cluster action. In Eq.~\eqref{twenty-one}, the cluster
self-energy has been removed from $\mathcal{G}$ to avoid double counting.

Then, the interaction on the cluster
\begin{equation}
\frac{U}{N_c} \sum_{\bf K,K^\prime,Q} c^\dagger_{\bf K+Q\uparrow} c_{{\bf K}\uparrow} c^\dagger_{\bf K^\prime-Q\downarrow}\,  c_{\bf K^\prime\downarrow}
\label{twenty-two}
\end{equation}
is linearized by introducing a discrete $\tau$-dependent Hubbard-Stratonovich
field on each $\tau$-slice and for each ${\bf K}$ point. In this way, the cluster problem is
transformed into a problem of non-interacting electrons coupled to $\tau$-dependent
Hubbard-Stratonovich fields.  Integrating out the bilinear fermion field and using
importance sampling to sum over the Hubbard-Stratonovich fields one evaluates the cluster
Green's function $G_c({\bf K}, \omega_n)$. From this, one evaluates the cluster self-energy
\begin{equation}
\Sigma_c({\bf K}, \omega_n) = \mathcal{G}^{-1} ({\bf K}, \omega_n)-G^{-1}_c ({\bf K}, \omega_n)\, .
\label{twenty-three}
\end{equation}
Then, using this new result for $\Sigma_c({\bf K},\omega_n)$ in Eq.~\ref{twenty}, these
steps are iterated to convergence.

Measurements of correlation functions and the 4-point vertex are made in the same manner as
for the determinantal Monte Carlo.  That is, after the Hubbard-Stratonovich field has been
introduced, one has a Wick's theorem for decomposing products of various time-ordered
operators.  However, in this case there is an additional coarse-graining of the Green's
function intermediate state legs\cite{ref:6,ref:21}.  Since one is using a determinantal
Monte Carlo method, there is also a sign problem for the doped Hubbard model.  However,
the self-consistent bath and the coarse-graining of the momentum space significantly
reduce this problem so that lower temperatures can be reached.\cite{ref:qmc}

\subsection*{The Density Matrix Renormalization Group}

\begin{figure}[htb]
\begin{center}
\includegraphics[width=10cm,angle=0]{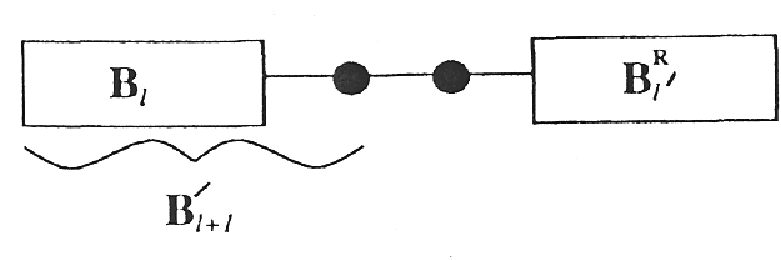}
\caption{The configuration of blocks used for the density matrix renormalization
group algorithm (White\protect\cite{ref:25}).}
\label{fig:3}
\end{center}
\end{figure}

The density-matrix-renormalization-group method was introduced by White.\cite{ref:24,ref:25}  Here, as
illustrated in Fig.~\ref{fig:3} for a one-dimensional chain of length $L$, the system under study is
separated into four pieces.  A block $B_\ell$ containing $\ell=L/2-1$ sites on the left, a
reflected $B^{\rm R}_{\ell^\prime}$ (right interchanged with left) block containing
$\ell^\prime=L/2-1$ sites on the right, and two
additional sites in the middle.  The left hand block $B_\ell$ and its near-neighbor site
are taken to be the system to be studied, while the block $B^{\rm R}_{\ell^\prime}$
plus its adjacent site are considered to be the ``environment''.  The entire system is
diagonalized using a Lanczos or Davidson algorithm to obtain the ground state eigenvalue
and eigenvector $\psi_\circ$. Then, one constructs a reduced density matrix from $\psi_\circ$
\begin{equation}
\rho_{ii^\prime}=\sum_j \psi^*_{\circ ij} \psi_{\circ i^\prime j}\, .
\label{twenty-four}
\end{equation}
Here, $\psi_{\circ ij}=\langle i|\langle j|\psi_\circ\rangle$ with $|i\rangle$ a basis state of the
$\ell+1$ block and $|j\rangle$ a basis state of the $\ell^\prime +1$ ``environment''
block.  Then the reduced density matrix $\rho_{ii^\prime}$ is diagonalized and $m$ eigenvectors, corresponding to the $m$
largest eigenvalues are kept.  The Hamiltonian $H_{\ell+1}$ for the left hand block and
its added site $B^\prime_{\ell+1}$ is now transformed to a reduced basis consisting of the $m$ leading
eigenstates of $\rho_{ii^\prime}$.  The right hand environment block $H^R_{\ell^\prime+1}$ is chosen to be a
reflection of the system block including the added site.  Finally, a superblock of size
$L+2$ is formed using $H_{\ell+1}, H^R_{\ell^\prime+1}$ and two new single sites.  Open boundary
conditions are used.
Typically, several hundred eigenstates of the reduced density matrix are kept and thus,
although the system has grown by two sites at each iteration, the number of total states
remains fixed and one is able to continue to diagonalize the superblock.

This infinite system method suffers because the groundstate wave function $\psi_\circ$ continues
to change as the lattice size increases.  This can lead to convergence problems.
Therefore, in practice, a related algorithm in which the length $L$ is fixed has been
developed. In this case, instead of trying to converge to the infinite system fixed point
under iteration, one has a variational convergence to the ground state of a finite system.
This finite chain algorithm is similar to the one we have discussed but in this case the total
length $L$ is kept fixed and the separation point between the system and the environment
is moved back and forth until convergence is achieved.\cite[5]{ref:25}  Following this, one
can consider scaling $L$ to infinity.

In a sense, the density-matrix-renormalization-group method is a cluster theory.  It
embeds a numerical renormalization procedure in a larger lattice in which an exact
diagonalization is carried out.  The division of the chain into the system and the
environment is similar in spirit to the embedded cluster and $\mathcal{G}^{-1}$. The use of the
reduced density matrix, corresponding to the groundstate, to carry out the basis truncation
provides an optimal focus on the low-lying states.

\begin{figure}[htb]
\begin{center}
\includegraphics[width=8cm,angle=0]{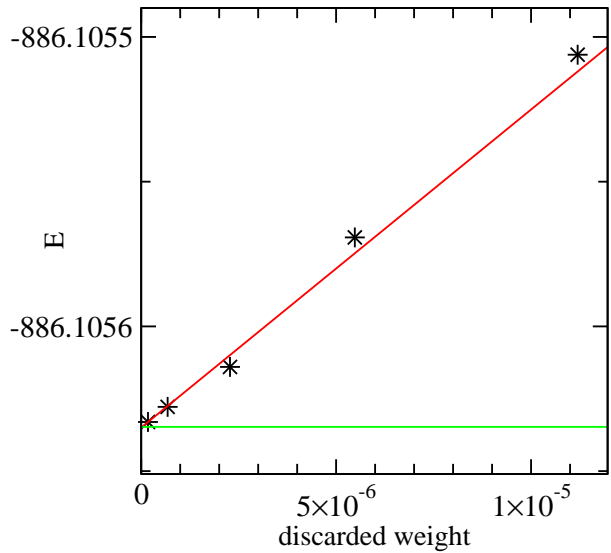}
\caption{DMRG results for the ground state energy of a 2000-site Heisenberg spin-one-half
chain versus the discarded weight $W_m$.  The exact Bethe ansatz energy is shown as the
line at the bottom of the figure. (S.R.~White)}
\label{fig:4}
\end{center}
\end{figure}

An important aspect of this approach is how rapidly the eigenvalues of the reduced density
matrix fall off.  This determines how many $m$ states one needs to obtain accurate results.
Unfortunately, for the study of $n$-leg Hubbard ladders, this fall-off becomes
significantly slower as $n$ increases and many more states must be kept.  In addition, it
appears that the behavior of the pairfield-pairfield correlation function is particularly
sensitive to the number of states $m$  that are kept. A measure of the errors associated
with the truncation in the number $m$ of density matrix eigenstates that are kept, is given
by the discarded weight
\begin{equation}
W_m=\sum^D_{i=m+1} w_i\, .
\label{twenty-five}
\end{equation}
Here, $D$ is the dimension of the density matrix and $w_i$ is its $i^{\rm th}$ eigenvalue.
A useful approach is to increase $m$ and extrapolate quantities to their values as
$W_m\to0$.  The error in the groundstate eigenvalue varies as $W_m$ and a typical
extrapolation is shown in Fig.~\ref{fig:4}.  For observables which do not commute with $H$, the
errors vary as $\sqrt{W_m}$.

\vspace{0.2in}

\section*{3. Properties of the 2D Hubbard Model}

As we have discussed, the particle-hole symmetry of the half-filled Hubbard model with a
near-neighbor hopping $t$ leads to an absence of the fermion sign problem.  In this case, the
determinantal Monte Carlo algorithm \cite{ref:22} provides a powerful numerical tool for studying the
low temperature properties of this model. In a seminal paper, Hirsch\cite{ref:11} presented numerical
evidence that the groundstate of the half-filled two-dimensional Hubbard model with a
near-neighbor hopping $t$ and a repulsive on-site interaction $U>0$ had long-range
antiferromagnetic order.  In this work, simulations on a set of L$\times$L lattices were
carried out.  For each lattice, simulations were run at successively lower temperatures and
extrapolated to the $T=0$ limit. Then, a finite-size scaling analysis was used to extrapolate
to the bulk $T=0$ limit.

This work set the standard for what one would like to do
in numerical studies of the Hubbard model.  Unfortunately, the fermion sign problem prevents
one from carrying out a similar determinantal Monte Carlo analysis for the doped case.
However, various other methods have been developed which provide information on the
doped Hubbard model.  Here, we will discuss results obtained from a dynamic cluster Monte
Carlo algorithm.\cite{ref:6} This method also provides a systematic approach to the bulk limit as
the cluster size increases.  As noted in Sec.~2, the dynamic cluster Monte Carlo still
suffers from the fermion
sign problem, although to much less of a degree than the standard determinantal Monte
Carlo. Maier \etal\cite{ref:16} have made the important step of studying the
doped system on a sequence of different-sized clusters ranging up to 26 sites in size.  Furthermore,
this work recognized the importance of cluster geometry and developed a Betts'-like\cite{ref:28} grading
scheme for determining which cluster geometries are the most useful in determining the
finite-size scaling extrapolation.

Following this, we review a density-matrix-renormalization-group (DMRG) study\cite{ref:29} of
a doped 6-leg Hubbard ladder in which the authors extrapolate their results to the limit of zero
discarded weight and to legs of infinite length.  This work provides evidence that static
stripes exist in the ground state
for large values of $U(U\simeq12t)$ but are absent at weaker coupling $(U\ltwid 3t)$.  We conclude this section
with a discussion of the pseudogap behavior which is observed in the lightly doped Hubbard model
when $U$ is of order the bandwidth.

\begin{figure}[htb]
\begin{center}
\includegraphics[width=10cm,angle=90]{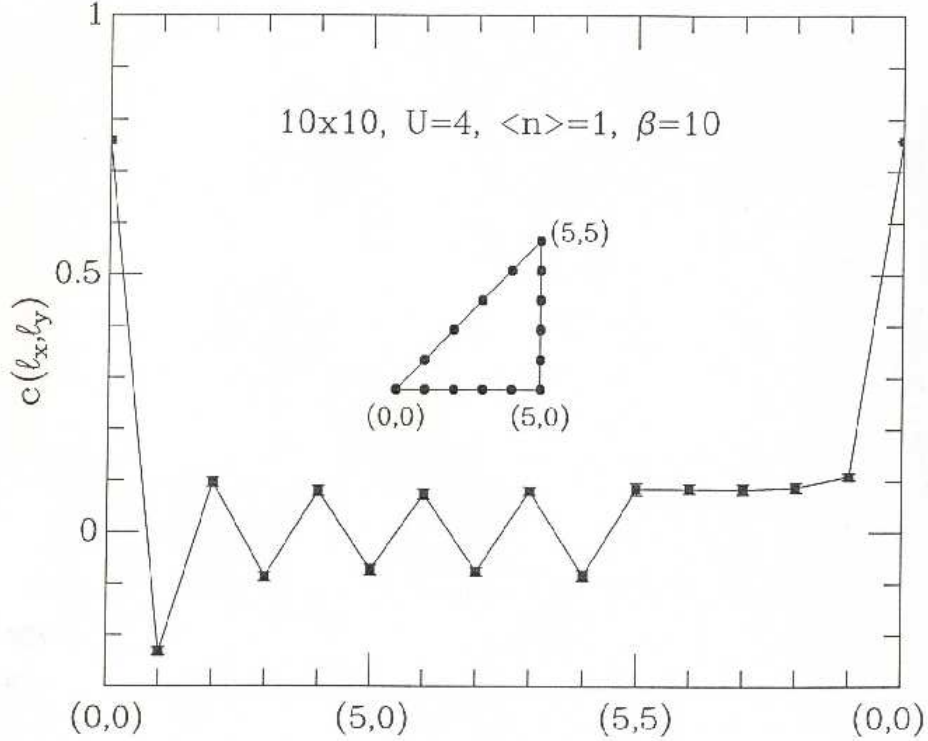}
%\vspace{7cm}
\caption{The equal-time magnetization-magnetization correlation function $C(\ell_x,\ell_y)$
on a $10\times10$ lattice with $U=4t$, $\langle n\rangle=1$ and $T=0.1t$.  The horizontal
axis traces out the triangular path on the lattice shown in the inset.  Strong antiferromagnetic
correlations are seen (Hirsch \protect\cite{ref:11}, Moreo \etal \protect\cite{ref:A2}).}
\label{fig:5}
\end{center}
\end{figure}

\subsection*{The Antiferromagnetic Phase}

Determinantal quantum Monte Carlo results for the equal-time magnetization-magnetization
correlation function
\begin{equation}
C(\ell) = \left\langle m^z_{i+\ell} m^z_i\right\rangle
\label{twenty-six}
\end{equation}
with $m^z_i=(n_{i\uparrow}-n_{i\downarrow})$ are plotted in Fig.~\ref{fig:5}. These results are for a
half-filled Hubbard model on a 10$\times$10 lattice at a temperature $T=0.1t$ with $U=4t$. At
this temperature, the antiferromagnetic correlation length exceeds the lattice size and the
cluster is essentially in its groundstate.  Strong antiferromagnetic correlations are clearly
visible in $C(\ell)$.

\begin{figure}[htb]
\begin{minipage}[t]{8cm}
%\begin{center}
%\vspace{7cm}
\centerline{\includegraphics[width=7.25cm,angle=90]{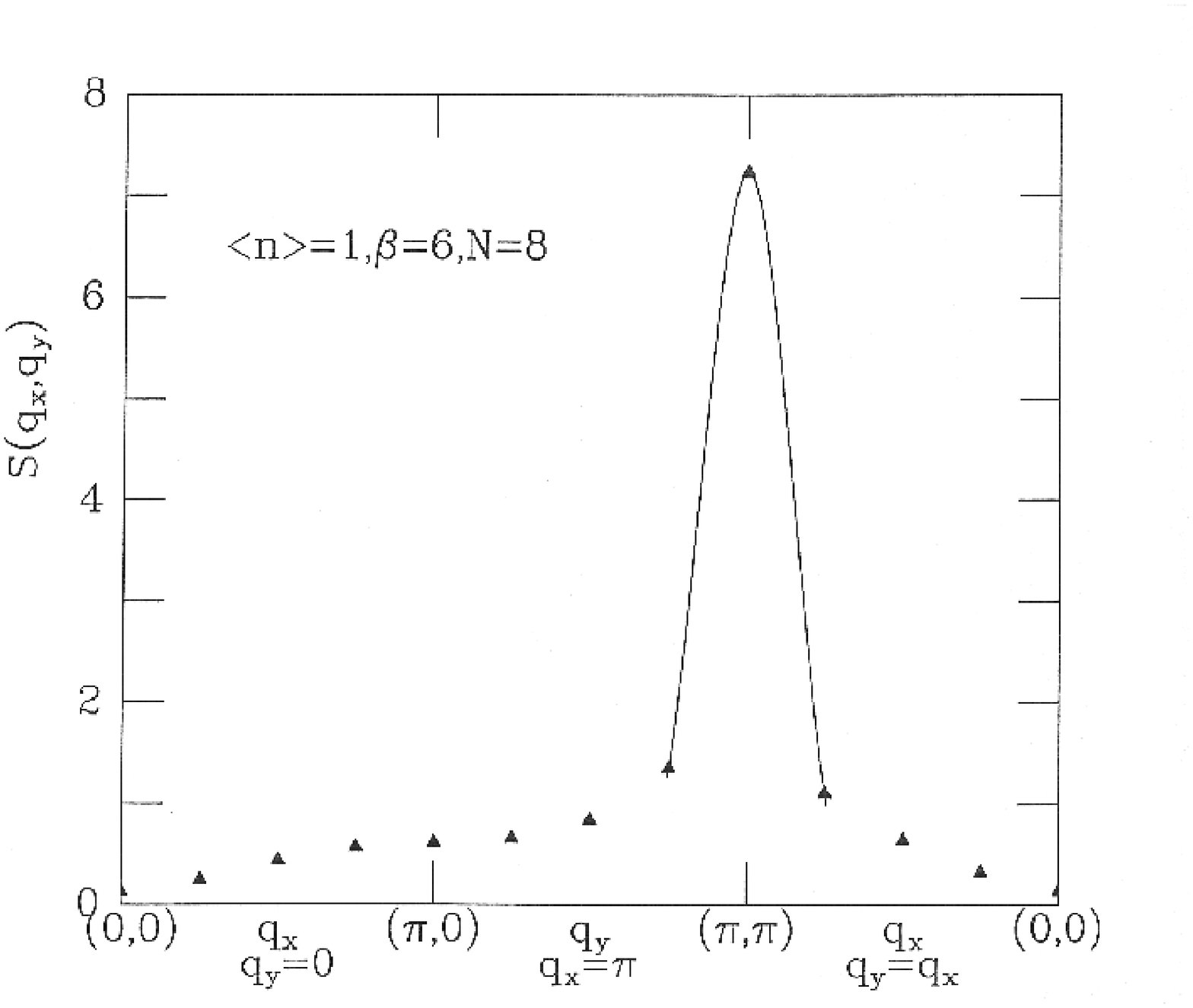}\phantom{x}}
\caption{$S(q_x,q_y)$ versus $q_x,q_y$ for $\langle n\rangle=1$, $U=4t$ and $T=.167t$.
The solid line is a fit to guide the eye (Hirsch \protect\cite{ref:11}, Moreo \etal \protect\cite{ref:A2}).}
\label{fig:6a}
%\end{center}
\vspace{0.5cm}
\end{minipage}
\hfill
\begin{minipage}[t]{8cm}
%\begin{center}
%\vspace{7cm}
\centerline{\includegraphics[width=7.25cm,angle=90]{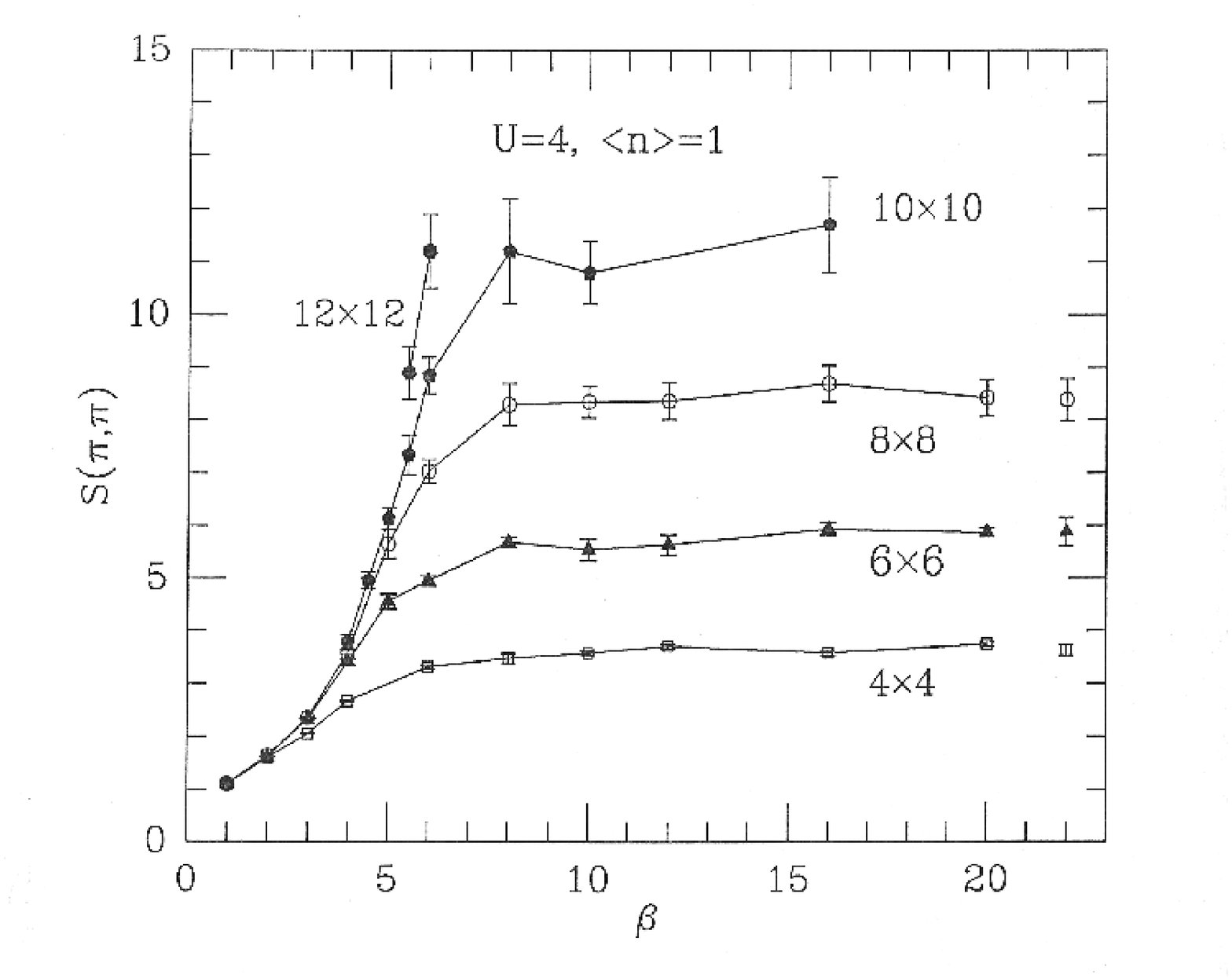}\phantom{x}}
\caption{The antiferromagnetic structure factor $S(\pi,\pi)$ for $\langle n\rangle=1$
and $U=4$ as a function of the inverse temperature $\beta$ for various lattice sizes.
$S(\pi,\pi)$ saturates when the coherence length exceeds the lattice size
(Hirsch \protect\cite{ref:11}, White \etal \protect\cite{ref:12}).}
\label{fig:6}
%\end{center}
\end{minipage}
\end{figure}

The magnetic structure factor, shown in Fig.~\ref{fig:6a}
\begin{equation}
S(q) = \frac{1}{N}\ \sum e^{-i{\bf q} \cdot {\bm\ell}} \left\langle m^z_{i+\ell}
\, m_i\right\rangle
\label{twenty-seven}
\end{equation}
has a peak at $q=(\pi, \pi)$ reflecting the antiferromagnetic correlations.  As shown in
Fig.~\ref{fig:6}, as the temperature is lowered $S(\pi,\pi)$ grows and then saturates when
the antiferromagnetic correlations extend across the lattice. If there is long-range
antiferromagnetic order in the groundstate, the saturated value of $S(\pi,\pi)$ will
scale with the number of lattice sites $N=L\times L$. Furthermore, based upon spin-wave
fluctuations,\cite{ref:30} one expects that the leading correction will vary as
$N^{\frac{1}{2}}$ so that
\begin{equation}
\lim_{N\to\infty} \frac{S(\pi, \pi)}{N} = \frac{\langle m_x\rangle^2}{3} +
\frac{A}{N^{\frac{1}{2}}}\, .
\label{twenty-eight}
\end{equation}

\begin{figure}[htb]
\begin{center}
%\vspace{7cm}
\includegraphics[width=10cm,angle=0]{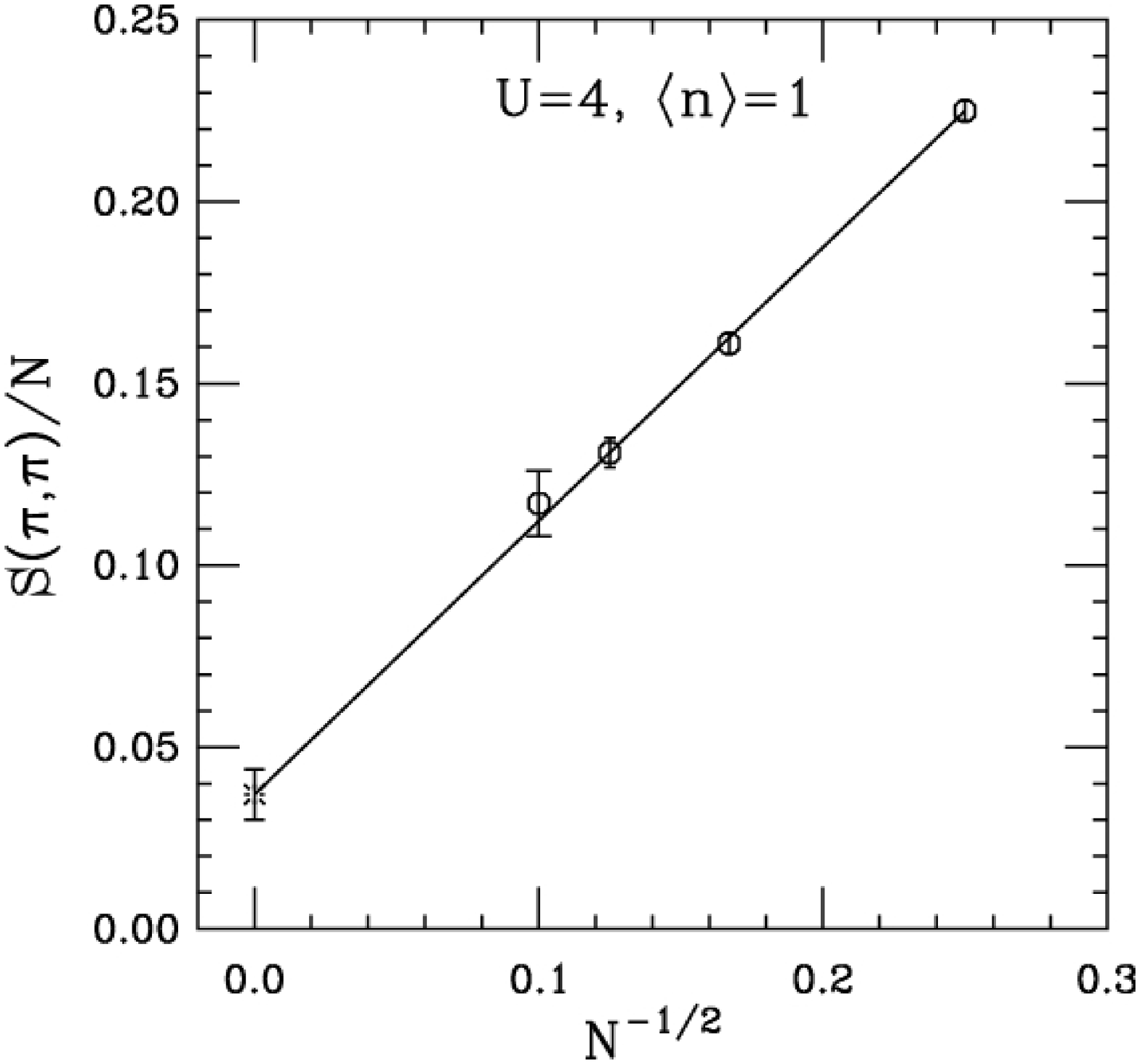}
\caption{The zero-temperature limit of $S(\pi,\pi)/N$ versus $1/N^{1/2}$.  The
results extrapolate to a finite value as $N\to\infty$ implying that there is long range antiferromagnetic
order in the groundstate of the infinite lattice (Hirsch \protect\cite{ref:11}, White \etal \protect\cite{ref:12}).}
\label{fig:7}
\end{center}
\end{figure}

Fig.~\ref{fig:7} shows $S(\pi, \pi)/N$ versus $N^{-\frac{1}{2}}$ for $U=4t$ and one sees that the
groundstate has long-range antiferromagnetic order.  In his original paper, Hirsch\cite{ref:11}
concluded that the groundstate of the half-filled 2D Hubbard model with a near-neighbor hopping
$t$ would have long-range antiferromagnetic order for $U>0$.

\subsection*{\boldmath$d_{x^2-y^2}$ Pairing}

\begin{figure}[htb]
\begin{minipage}[b]{8cm}
%\begin{center}
\centerline{\phantom{xxxxxxxxx}\includegraphics[width=8.80cm,angle=-90]{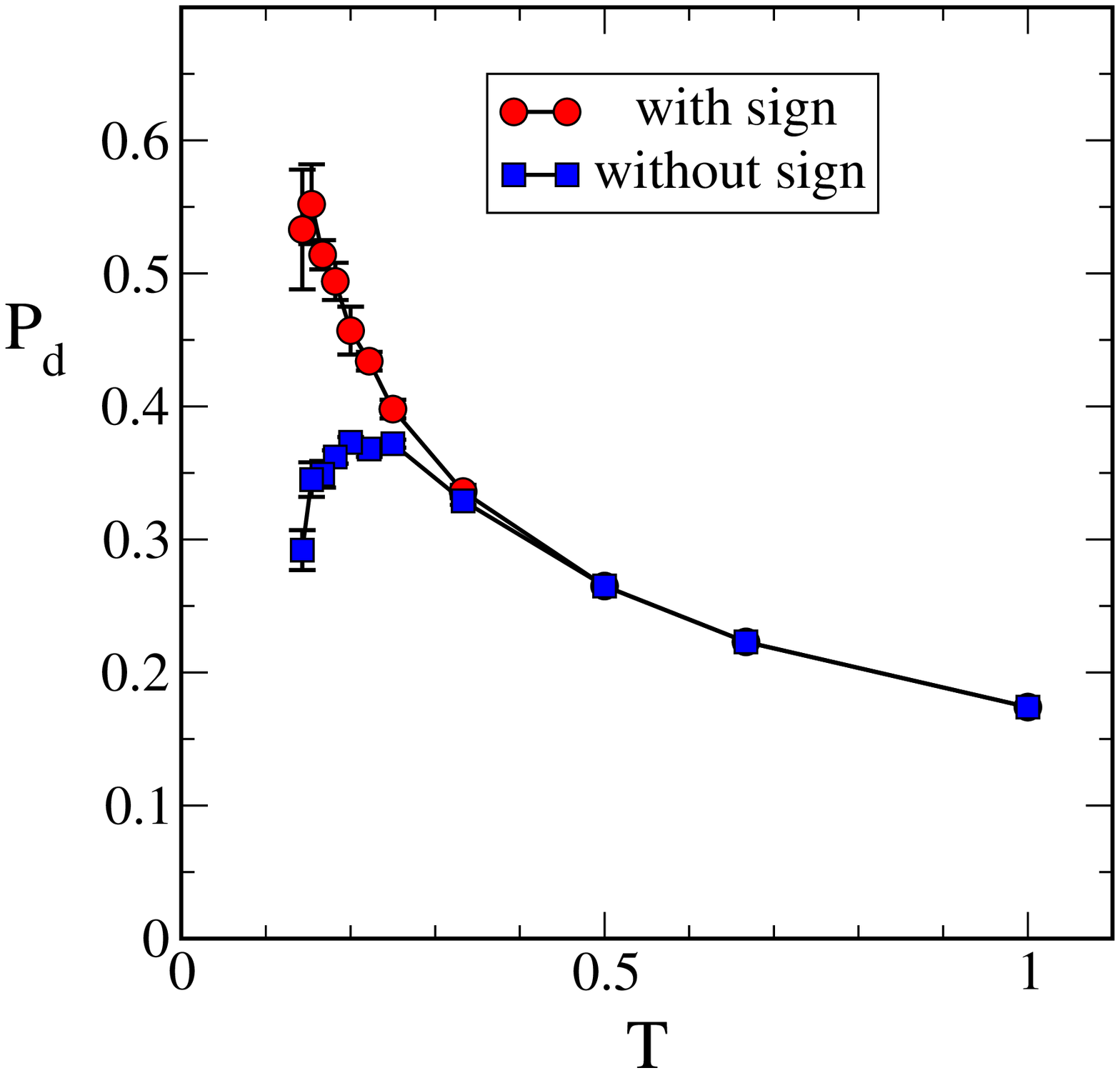}}
\vspace{-0.1cm}
\caption{The d-wave pairfield susceptibility $P_d(T)$ (red circles) for a $4\times4$
lattice with $U=4t$ and $\langle n\rangle=0.875$ versus temperature $T$ measured in
units of the hopping $t$.  The (blue squares) show the erroneous result that is found if
the fermion sign is ignored. (Loh \etal\protect\cite{ref:27})}
\label{fig:8}
%\end{center}
\vspace{0.980cm}
\end{minipage}
\hfill
\begin{minipage}[b]{8cm}
%\begin{center}
\centerline{\includegraphics[width=7.5cm,angle=0]{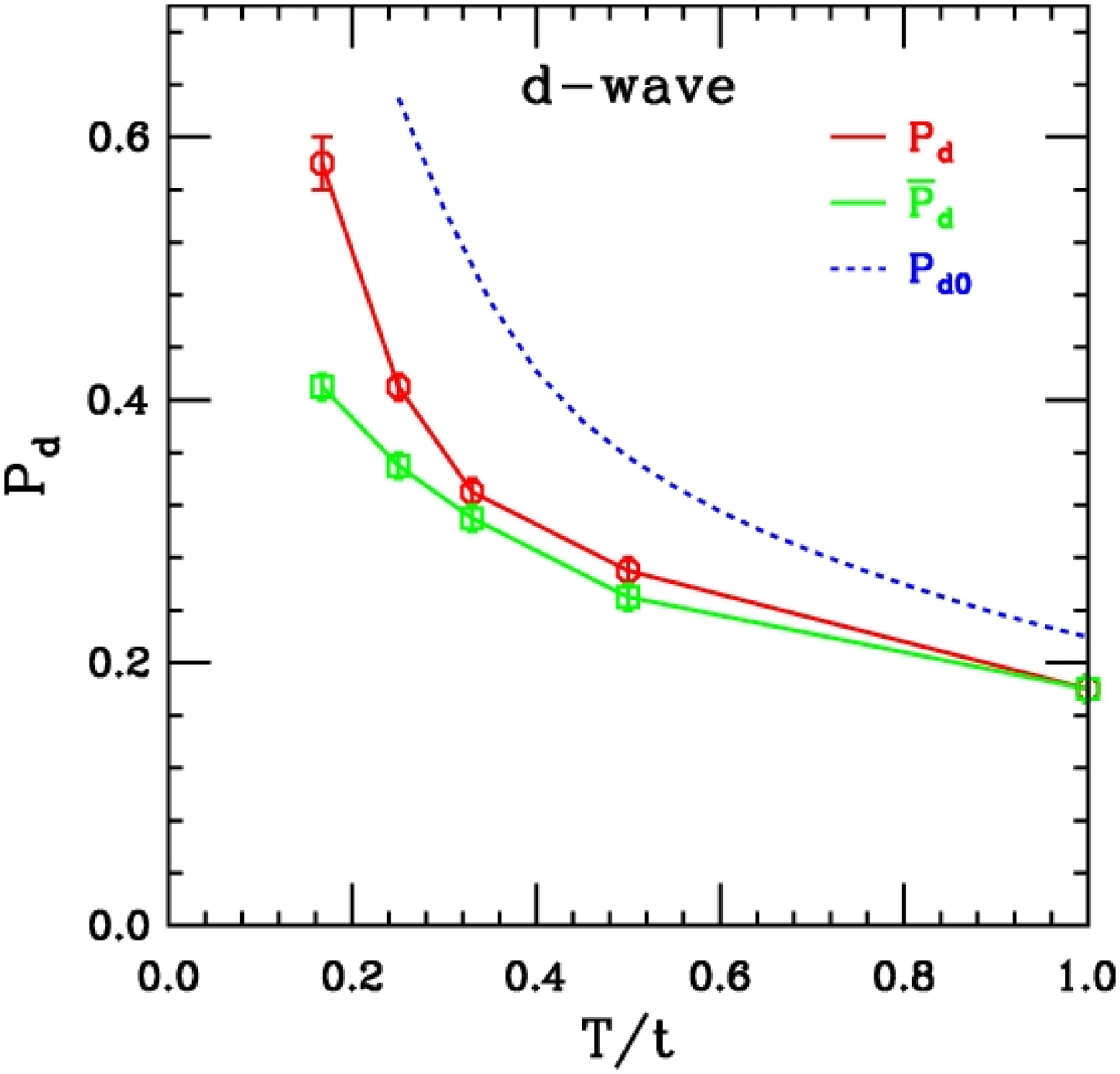}\phantom{xx}}
\caption{The d-wave pair-field susceptibility $P_d(T)$ is shown as the open (red) circles.
The open (green) squares show results for the ``noninteracting" pair-field susceptibility
$\overline{P}_d(T)$ calculated using dressed single-particle Green's functions,
Eq.~\protect\ref{thirty-one}, while the dashed (blue) curve is the noninteracting susceptibility
$P_{d\circ}$ calculated with the bare Green's functions. (White \etal\protect\cite{ref:12})}
\label{fig:9}
%\end{center}
\end{minipage}
\end{figure}

The structure of the pairing correlations in the doped 2D Hubbard model were initially
studied using the determinantal Monte Carlo method. The d-wave pairfield susceptibility
\begin{equation}
P_d=\int^\beta_0 d\tau\, \langle \Delta_d (\tau)\, \Delta^\dagger_d(0)\rangle
\label{twenty-nine}
\end{equation}
with
\begin{equation}
\Delta^\dagger_d=\frac{1}{2\sqrt{N}}\ \sum_{\ell, \delta} (-1)^\delta
c^\dagger_{\ell\uparrow} c^\dagger_{\ell+\delta\downarrow}
\label{thirty}
\end{equation}
was calculated.  Here $\delta$ sums over the four near-neighbor sites of $\ell$ and
$(-1)^\delta$ gives the $+-+-$ sign alteration characteristic of d-wave pairing. The doped
Hubbard model has a fermion sign problem, so that the Hubbard-Stratonovich fields must be
generated according to the probability distribution $P_\|(S)$ given by Eq.~\eqref{sixteen}.
In this case, it is essential to include the sign factor $s$ in the evaluation of observables.  The (red) circles in
Fig.~\ref{fig:8} show results\cite{ref:27} for $P_d(T)$ obtained on a 4$\times$4 lattice with $\langle
n\rangle=0.875$ and $U=4t$.  If the sign $s$ is not included, one obtains the (blue) squares.
The neglect of this sign in early work\cite{ref:31} left the false impression that the Hubbard model did
not support $d_{x^2-y^2}$ pairing.

As seen, when the sign is included, the d-wave pairfield susceptibility
increases as the temperature is lowered. However, over the temperature range accessible to
the determinantal Monte Carlo, it remains smaller than the $U=0$ result $P_{d\circ}$,
shown as the (blue) dashed
line in Fig.~\ref{fig:9}.  In Ref.~\cite{ref:12}, it was argued that this behavior was due to the renormalization
of the single particle spectral weight and that the significant feature to note was that
$P_d(T)$ was enhanced over
\begin{equation}
\overline{P}_d (T) = \frac{T}{N} \sum_{pn} G(p)\, G(-p)\, (\cos p_x-\cos p_y)^2\, .
\label{thirty-one}
\end{equation}
Here, $G(p)$ is the dressed single particle Green's function determined from the Monte Carlo
simulation and $\overline{P}_d$ corresponds to the contribution of a pair of dressed but non-interacting
holes.  The fact that $\overline{P}_d(T)$ lays below $P_d(T)$ implies that there is an attractive
$d_{x^2-y^2}$-pairing interaction between the holes.  $\overline{P}_d(T)$ is shown as the
(green) curve labeled with open squares in Fig.~\ref{fig:9}.

%\begin{figure}[htb]
%\begin{center}
%\includegraphics[width=10cm,angle=90]{filename.ext}
%\caption{Diagrammatic representation of the pairfield susceptibility in terms of dressed
%Green's functions $G$ (solid lines) and the reducible particle-particle vertex $\Gamma^{pp}$.}
%\label{fig:10}
%\end{center}
%\end{figure}
%
%An alternative way to look at this begins with the representation of $P_d$
%illustrated in Fig.~\ref{fig:10}.  Here, the solid lines are dressed Green's functions and
%$\Gamma^{\rm pp}$ is the two-particle irreducible vertex. If one were to approximate the
%effective pairing interaction $\Gamma^{\rm pp} (p|p^\prime)$ by a simple separable form
%\begin{equation}
%\Gamma^{\rm pp}(p|p^\prime)=-V_d (\cos p_x-\cos p_y)\, (\cos p^\prime_x-\cos p^\prime_y)
%\label{thirty-two}
%\end{equation}
%then
%\begin{equation}
%P_d\cong \frac{\overline{P}_d}{1-V_d\overline{P}_d}\, .
%\label{thirty-three}
%\end{equation}
%From this one has
%\begin{equation}
%V_d=\frac{1}{\overline{P}_d}-\frac{1}{P_d}\, .
%\label{thirty-four}
%\end{equation}
%From the Monte Carlo results for $\overline{P}_d$ and $P_d$ shown in Fig.~\ref{fig:9}, one sees that
%there is an attractive d-wave pairing interaction and that its strength $V_d$ increases as the
%temperature is lowered. We expect
%that it saturates at temperatures below the exchange interaction, but because of the fermion
%sign problem, the determinantal Monte Carlo is unable to explore this region.

\begin{figure}[htb]
\begin{center}
\includegraphics[width=\textwidth]{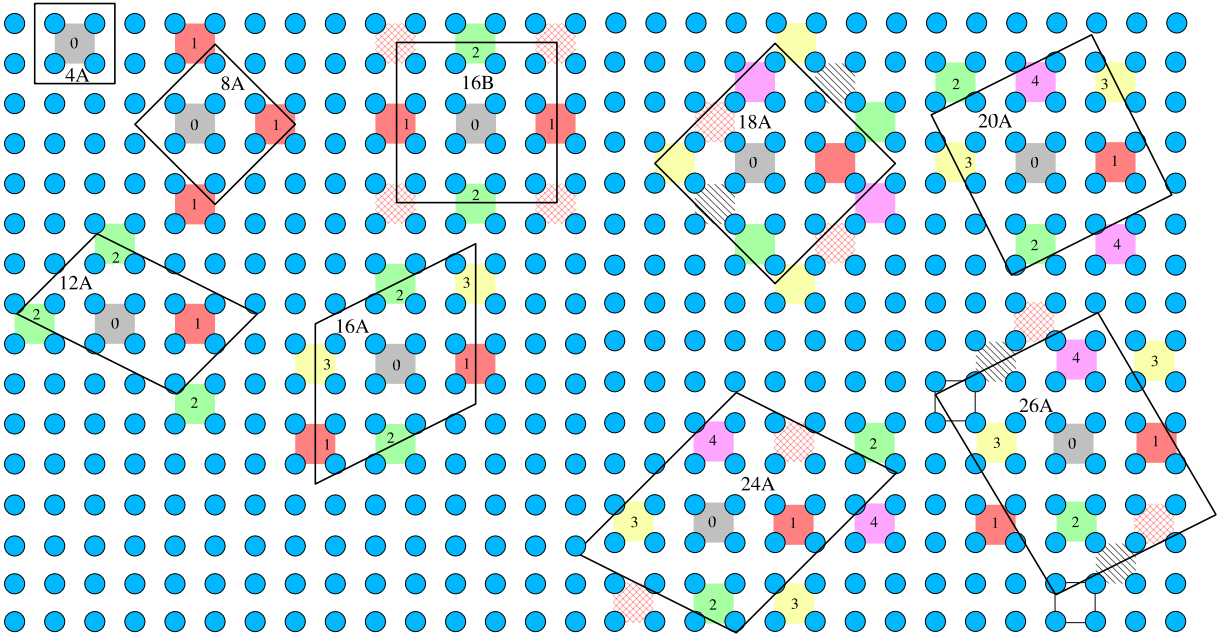}
\caption{Cluster sizes and geometries used by Maier \etal\cite{ref:16} in their study of the
d-wave pair-field susceptibility.  The shaded squares represent independent d-wave plaquettes
within the clusters.  In small clusters, the number of neighboring d-wave plaquettes $Z_d$
listed in Table~1 is smaller than 4, i.e., than that for an infinite lattice. 
(Maier \etal\protect\cite{ref:16})}
\label{fig:11}
\end{center}
\end{figure}

In order to determine what happens at lower temperatures, Maier \etal\cite{ref:16} have
determined $P_d(T)$ using a dynamic cluster approximation.  In a systematic study, they
provided evidence that the doped Hubbard model contained a $d_{x^2-y^2}$ pairing phase.
In this work, the authors adapted a cluster selection criteria originally introduced by
Betts \etal\cite{ref:28} in a numerical study of the 2D Heisenberg model.  For the
Heisenberg model, Betts \etal\cite{ref:28} showed that an important selection criteria
for a cluster was the completeness of the
``allowed neighbor shells'' compared to an infinite lattice.  They found that a finite-size
scaling analysis was greatly improved when clusters with the most complete shells were
selected.  For a d-wave order parameter, Maier \etal noted that one needs to take into
account the non-local 4-site plaquette structure of the order parameter in applying this
criteria.  As illustrated in Fig.~\ref{fig:11}, the 4-site cluster encloses just one d-wave plaquette.
Denoting the number of independent near-neighbor plaquettes on a given cluster by $Z_d$,
the 4-site cluster has no near-neighbors so that $Z_d$=0. In this case the embedding action
does not contain any pair field fluctuations and hence $T_c$ is over estimated.
Alternatively, the 8A cluster has space for one more 4-site plaquette $(Z_d=1)$ and the
same neighboring plaquette is adjacent to its partner on all four sides. In this case the
phase fluctuations are over estimated and $T_c$ is suppressed.
For the 16B cluster, one has $Z_d=2$ while $Z_d=3$ for the oblique 16A cluster. Thus, one
expects that the pairing correlations for the 16B cluster will be suppressed relative to
those for the 16A cluster. The number of independent neighboring d-wave plaquettes $Z_d$
for the clusters shown in Fig.~\ref{fig:11} are listed in Table~1.

\begin{table}[h]
\begin{center}
\begin{tabular}{ccl}
\hline\hline
Cluster&$Z_d$&\hphantom{00}$T_c/t$\\ \hline
4&0 (MF)&0.056\\
\hphantom{2}8A&1&$-$0.006\\
18A&1&$-$0.022\\
12A&2&\hphantom{$-$}0.016\\
16B&2&\hphantom{$-$}0.015\\
16A&3&\hphantom{$-$}0.025\\
20A&4&\hphantom{$-$}0.022\\
24A&4&\hphantom{$-$}0.020\\
26A&4&\hphantom{$-$}0.023
\end{tabular}
\caption{Number of independent neighboring d-wave plaquettes $Z_d$ and the value
$T_c$ obtained from a linear fit of the pair-field susceptibility in Fig.~\protect\ref{fig:12}
(Maier \etal \protect\cite{ref:16}).}
\end{center}
\end{table}

\begin{figure}[htb]
\begin{center}
\includegraphics[width=10cm,angle=0]{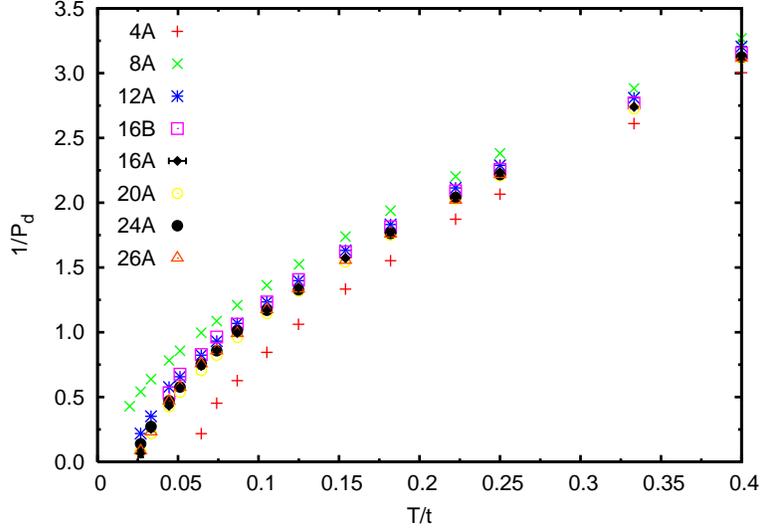}
\caption{The inverse of the d-wave pair-field susceptibility is plotted versus $T/t$ for
various clusters.  Here $U=4t$ and $\langle n\rangle=0.9$. (Maier \etal\protect\cite{ref:16})}
\label{fig:12}
\end{center}
\end{figure}

Results for the inverse of the pair field susceptibility versus $T$ for $U=4t$ and $\langle
n\rangle=0.9$ are shown in Fig.~\ref{fig:12}.  As expected, the 4-site cluster results over
estimate $T_c$ and the results for the 8A and 18A clusters do not give a positive value for
$T_c$. However, successive $Z_d=4$ results  on larger
lattices fall nearly on the same curve.  These results suggest that the 2D Hubbard model with
$U=4t$ and $\langle n\rangle =0.9$ has a $d_{x^2-y^2}$ pairing phase.
The dynamic cluster approximation leads to a
mean field behavior close to $T_c$.\cite{ref:?} Values of $T_c$ obtained using a mean field
linear fit of the low temperature data for the various clusters are listed in Table~1.

\begin{figure}[htb]
\begin{center}
\includegraphics[width=10cm,angle=0]{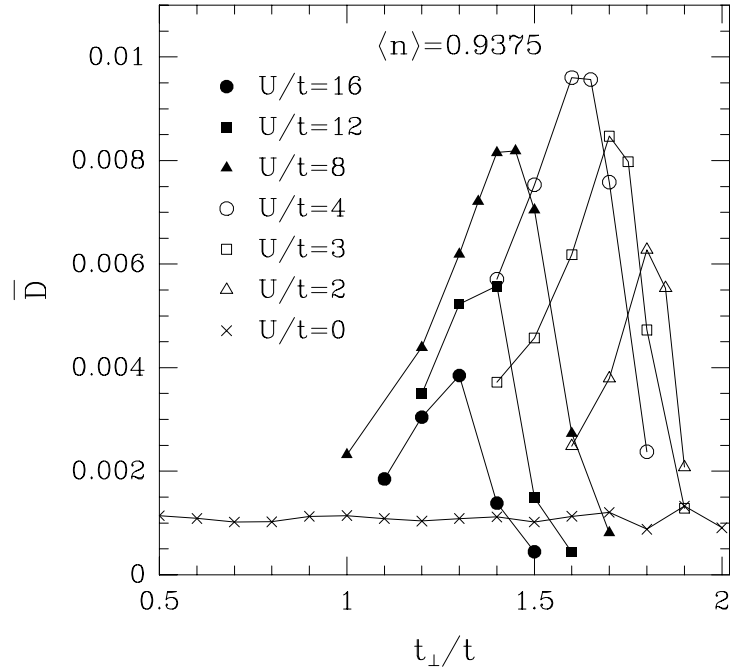}
\caption{$\bar D$ versus $t_\perp/t$ for various values of $U/t$ at a filling $\langle n\rangle=
0.9375$ (Noack \etal\protect\cite{ref:C1}).}
\label{fig:12a}
\end{center}
\end{figure}

If $T_c\simeq.02t$ and we take $t=0.2$~eV, this gives $T_c\sim50K$.  We believe
that $T_c$ will increase with $U$, reaching a maximum when $U$ is of order the bandwidth.  In addition,
we expect that the transition temperature is sensitive to the one-electron tight binding
parameters.  An example which illustrates this is known from density matrix renormalization-group
calculations for the 2-leg Hubbard ladder.\cite{ref:C1}  Figure~\ref{fig:12a} shows an average
of the rung-rung pairfield correlations
\begin{equation}
\overline{D}=\sum_\ell\langle\Delta(i+\ell)\Delta^\dagger(i)\rangle
\label{eq:D1}
\end{equation}
for a $2\times16$ Hubbard ladder versus the ratio of the rung to leg hopping parameters
$t_\perp/t$.  Here
\begin{equation}
\Delta^\dagger(i)=c^\dagger_{i1\uparrow}c^\dagger_{i2\downarrow}-
c^\dagger_{i1\downarrow}c^\dagger_{i2\uparrow}
\label{eq:C1}
\end{equation}
creates a pair on the $i^{\rm th}$ rung of the ladder.  The pairing, as measured by $\overline{D}$
exhibits a maximum at a value of $t_\perp/t$ when the minimum of the antibonding band at
$k_x=0$ and the maximum of the bonding band at $k_x=\pi$ approach the fermi surface.
For the half-filled noninteracting system, this would occur when $t_\perp/t=2$.
The doping and the interaction $U$ leads to a reduction of this ratio and to a flattening
of the dispersion which further enhances the single particle spectral weight near the
fermi energy.  If one considers the antibonding band to have $k_y=\pi$ and the bonding band
to have $k_y=0$, then this behavior is similar to increasing the single-particle spectral
weight near $(0,\pi)$ and $(\pi,0)$ in the 2D Hubbard model.  One also sees that the largest
peak in $\overline{D}$ occurs when $U$ is of order the bandwidth.

Having argued that the bandstructure plays a key role in determining $T_c$, it is important
to note that this raises a puzzle.  State of the art LDA calculations, as Andersen and
coworkers have shown,\cite{ref:C2} can be folded down to give material specific near-neighbor
$t$, next-near-neighbor $t^\prime$, etc. hopping parameters.  For the one-band Hubbard model
one would then have for the one-electron energy
\begin{equation}
{\cal E}_k=-2t(\cos k_x+\cos k_y)-4t^\prime\cos k_x\cos k_y-\dots
\label{eq:D2}
\end{equation}
From an analysis of a large number of hole-doped cuprates, it was found that $T_c$ is
correlated with the range of the intra-layer hopping.\cite{ref:C3} For the one-band Hubbard
model that we have discussed, this analysis implies that $T_c$ should increase as $t^\prime/t$ becomes
more negative.  The opposite trend is seen in both dynamic cluster\cite{ref:C4}
and density matrix renormalization-group calculations.\cite{ref:B1,ref:C6}  However,
a projected fermion calculation\cite{ref:ppar} finds that $t^\prime$ enters the effective
interaction and can lead to an increase in $T_c$ which is consistent with the conclusions
of Ref.~\cite{ref:C3}.  The resolution of this puzzle represents an important open problem.

\subsection*{Stripes}

In a DMRG study of 7r$\times$6 Hubbard ladders with 4r holes, Hager \etal\cite{ref:29} found that the
ground state was striped for strong coupling values of $U(U=12t)$. Using a systematic
extrapolation they gave evidence that such stripes exist in infinitely long 6-leg ladders.  These
studies also found that for small values of $U(U=3t)$ there were no stripes in the ground
state.  This work extended earlier work\cite{ref:WS} on a 7$\times$6 system with 4 holes
which found that a well-defined stripe formed for $U/t\sim 8$ to 12. The absence of stripes
for weak coupling is consistent with the fact that weak coupling renormalization group
studies of the Hubbard model find no evidence of a stripe instability.\cite{ref:RNG1,ref:34}

\begin{figure}[htb]
\begin{center}
\includegraphics[width=10cm,angle=0]{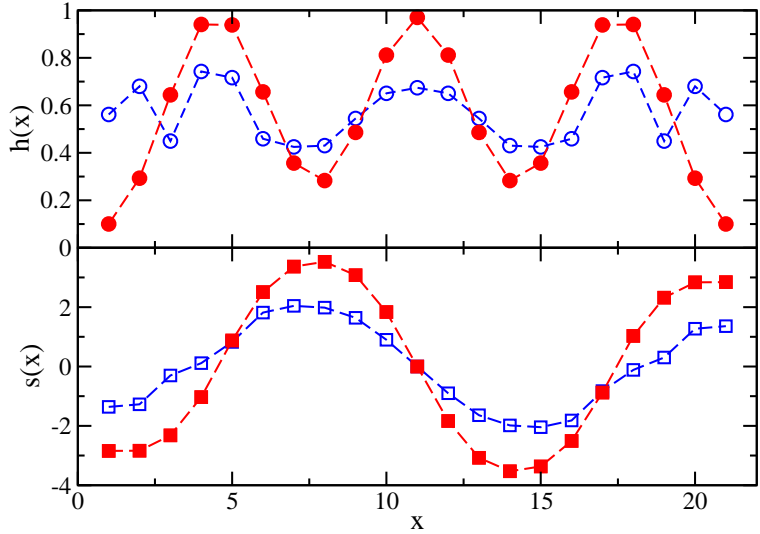}
\caption{The hole $h(x)$ (circle) and staggered spin $s(x)$
(square) densities in the leg x-direction are plotted for a $21\times6$ ladder with 12 holes
for $U=12t$ (solid symbols) and $U=3t$ (open symbols). (Hager \etal\protect\cite{ref:29})}
\label{fig:13}
\end{center}
\end{figure}

Using the DMRG technique, the ground state expectation values of the hole-density
\begin{equation}
h(x) = \sum^6_{y=1} (1-\langle n(x, y)\rangle)
\label{thirty-five}
\end{equation}
and the staggered spin density
\begin{equation}
s(x)=\sum^6_{y=1} (-1)^{x+y}\langle n_\uparrow (x, y)-n_\downarrow (x, y)\rangle
\label{thirty-six}
\end{equation}
were evaluated for 7r$\times$6 ladders with 4r holes.  Periodic boundary conditions were
used for the 6-site direction and open boundaries were used in the leg direction. Results
for the hole $h(x)$ and spin $s(x)$ densities for a
21$\times$6 ladder doped with 12 holes are shown in Fig.~\ref{fig:13}. One sees from the modulation of
the hole density along the leg direction that three stripes have formed. These stripes, each
associated with 4 holes, run around the 6-site cylinder. In earlier t-J studies,\cite{ref:10} a preferred
stripe density of half-filling was found and we believe that the 2/3 filling seen in Fig.~\ref{fig:13}
is a consequence of the restriction to 6 legs.  Just as in the t-J ladder calculations, the
staggered spin density undergoes a $\pi$-phase shift where the hole density is maximal.
The finite staggered spin density is an artifact of the DMRG procedure in which no spin
symmetry was imposed. This, along with the open boundary conditions which break the
translational symmetry allow the charge and spin density structures to appear in the
$h(x)$ and $s(x)$ expectation values.

While stripe-like structures are seen in Fig.~\ref{fig:13} for both $U/t=12$ and $U/t=3$, the
amplitude of the charge density modulations for $U/t=3$ are both smaller and less regular
than the $U/t=12$ results. As discussed in Sec.~2, DMRG results for operators which are
non-diagonal in the energy basis are expected to deviate from their exact values by the
square root of the discarded weight $\sqrt{W_m}$ as the number of basis states is
increased.  Thus, to determine whether there are stripes in the ground state of an infinite
ladder, Hager \etal\cite{ref:29} extrapolated their results for a set of 7r$\times$6 ladders to
$W_m\to 0$ and then took $R=7r\to\infty$.
\begin{figure}[htb]
\begin{center}
\includegraphics[width=10cm,angle=0]{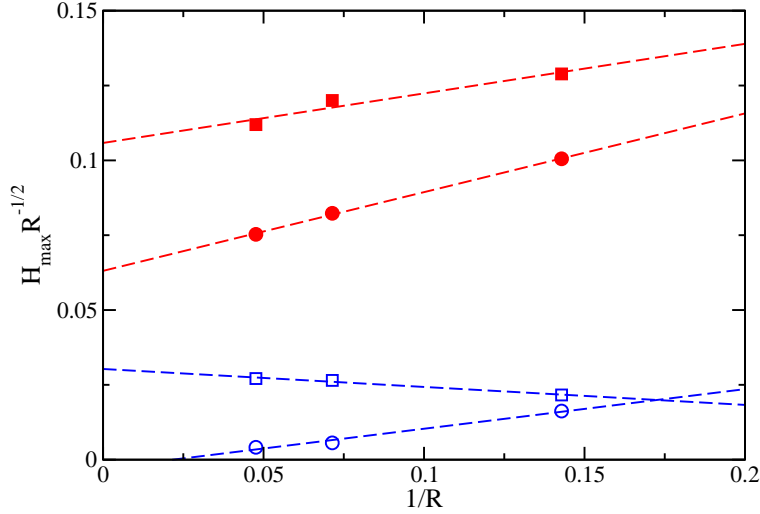}
\caption{The amplitude of the power spectrum component $|H(k^*_x)|/\sqrt R$ for the hole
density modulation.  Results for a fixed number $(6000\le m\le8000)$ of density-matrix
eigenstates (squares) and results extrapolated to the limit $W_m\to0$ (circles) are shown
as a function of the inverse ladder length $1/R$ for $U=12t$ (solid symbols) and $U=3t$
(open symbols).  The dashed lines are linear fits. (Hager \etal\protect\cite{ref:29})}
\label{fig:14}
\end{center}
\end{figure}
They did this for the wave-vector power spectrum of the charge density
\begin{equation}
H^2=\sum_{k_x} |\, H(k_x)\, |^2
\label{thirty-seven}
\end{equation}
with
\begin{equation}
H(k_x)=\sqrt{\frac{2}{R+1}}\ \sum_x \sin(k_x x)\, \langle h(x)\rangle\, .
\label{thirty-eight}
\end{equation}
For a ladder with a periodic array of stripes separated by 7 sites, the maximum contribution
to $H^2$ is associated with the wave vector
\begin{equation}
\frac{k^*_x}{\pi}=\frac{2r+1}{R+1} \to \frac{2}{7}
\label{forty}
\end{equation}
and
\begin{equation}
H_{\rm max} = |\, H(k^*_x)\, | \propto \sqrt{R}\ h_0
\label{thirty-nine}
\end{equation}
as $R$ goes to infinity. In Fig.~\ref{fig:14}, the amplitude $H_{\rm max} (k^*_x)/\sqrt{R}$ is plotted
for $U/t=12$ and $U/t=3$ versus the inverse of the ladder length $R^{-1}$. The solid
squares show the results when a fixed number $(6000 \leq m \leq 8000)$ of density-matrix
eigenstates are retained. The solid circles are the extrapolated $W_m\to0\, (m\to\infty)$
results for $U/t=12$. Similar results are shown using open symbols for $U/t=3$. When the
$W_m\to 0$ results are then extrapolated to $R\to\infty$, one sees clear evidence for
stripes when $U/t=12$ and an absence of stripes for $U/t=3$.  Note the importance of the $W_m\to0$
extrapolation in determining the absence of stripes for $U/t=3$.

\subsection*{The Pseudogap}

Besides the antiferromagnetic, d-wave pairing, and striped phases, the
cuprates exhibit a normal state pseudogap below a characteristic temperature $T^*$
when they are underdoped.  This pseudogap manifests itself in a variety of ways.\cite{ref:35}
There is a decrease in the Knight shift, reflecting a decrease in the
magnetic susceptibility.\cite{ref:All}  This was interpreted in terms of the opening of a pseudogap
in the spin degrees of freedom.  Observations of a similar suppression in the tunneling
density of states,\cite{ref:Tun} the c-axis optical conductivity\cite{ref:IR} and the specific heat\cite{ref:sp}
made it clear that there was a pseudogap in both the spin and charge degrees of freedom.
ARPES studies show that in the hole-doped materials, a pseudogap opens
near the $(\pi, 0)$ antinodal regions while in the electron-doped materials,
at the lowest dopings, it opens along the nodal direction near
$(\frac{\pi}{2}, \frac{\pi}{2})$.\cite{ref:36}  The pseudogap
appears in the underdoped region of the phase diagram and weakens as optimal doping is
approached.  If the Hubbard model is to contain the essential physics of the
cuprates, it should exhibit the pseudogap phenomenon.

Before looking for evidence of pseudogap behavior in the doped Hubbard model, it is useful to
first look at the structure of the single particle spectral weight for the half-filled
Hubbard model.  An important paper on this was that of Preuss \etal\cite{ref:A4} Here, determinantal
Monte Carlo calculations of the finite temperature single particle Green's function $G(k,
\tau)$ were carried out on an 8$\times$8 periodic lattice.  The spectral weight
\begin{equation}
A(k, \omega) = - \frac{1}{\pi}\ Im\, G(k, i\omega_n\to\omega + i\delta)
\label{forty-one}
\end{equation}
\begin{figure}[htb]
\begin{center}
\includegraphics[width=10cm,angle=0]{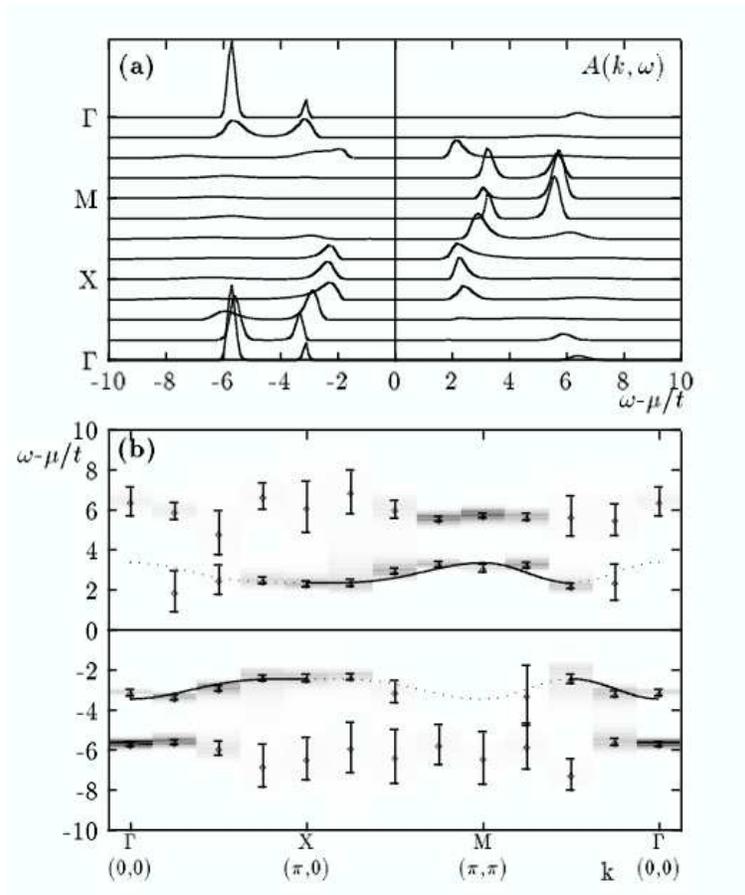}
\caption{Single-particle spectral weight $A(k,\omega)$ for an $8\times8$ Hubbard model
at half-filling $\langle n\rangle=1$ with $U=8t$ and $T=0.1t$.  (a) $A(k,\omega)$
versus $\omega$ for different values of $k$ and (b) $\omega$ versus $k$ plotted as a ``band-structure" where
sizable structure in $A(k,\omega)$ is represented by the strongly shaded regions and peaks
by error bars. (Preuss \etal\protect\cite{ref:A4})}
\label{fig:15}
\end{center}
\end{figure}
was then determined using a numerical maximum entropy continuation.  Results for the
half-filled case with $U=8t$ and $T=0.1t$ are shown in Fig.~\ref{fig:15}a. Here,
$A(k, \omega)$ is plotted versus $\omega$ for various $k$ values in
the Brillouin zone.  Figure~\ref{fig:15}b summarizes these results
using a standard ``band structure'' $\omega$ versus $k$ plot in which the dark areas signify
a large spectral weight.
This work and related studies \cite{ref:MH} showed that when $U$ was of order the bandwidth or
larger, there were four bands consisting of two incoherent upper and lower Hubbard bands and two
quasiparticle-like, narrow bands nearer $\omega=0$.  The inner bands were found to have a dispersion
set by $J\cong 4t^2/U$ while the outer, upper and lower Hubbard bands, appear as an essentially
dispersionless incoherent background.

\begin{figure}[htb]
%\begin{center}
\centerline{\includegraphics[width=14cm]{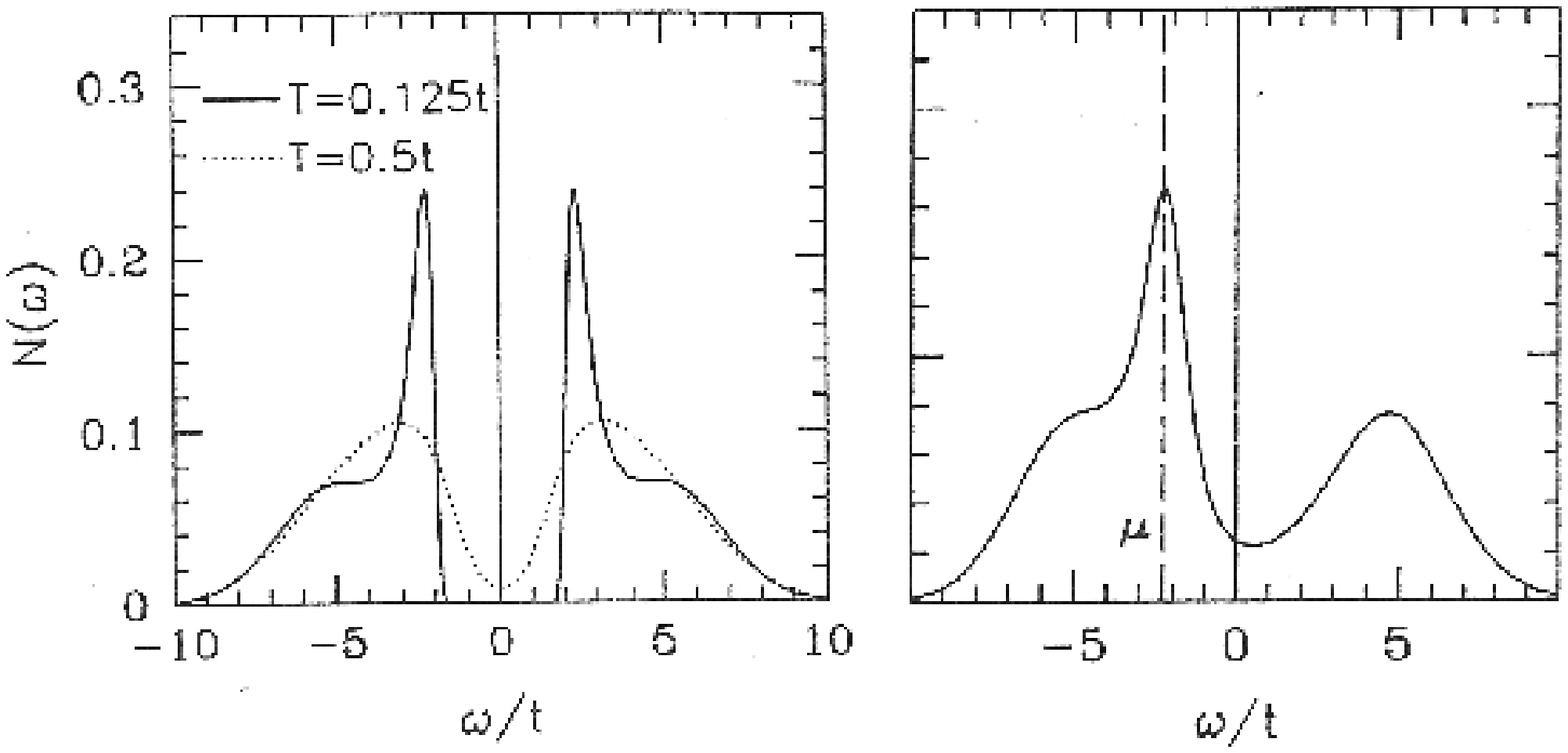}\phantom{xxx}}
\caption{On the left, the single particle density of states $N(\omega)$ versus $\omega$
for $U=8t$ and $\langle n\rangle=1$. On the right, $N(\omega)$
for the hole doped $\langle n\rangle=0.875$, $U=8t$ case at $T=0.33t$ (Scalapino \protect\cite{ref:Scal}).}
\label{fig:16}
%\end{center}
\end{figure}

\begin{figure}[htb]
\begin{center}
\includegraphics[width=10cm,angle=0]{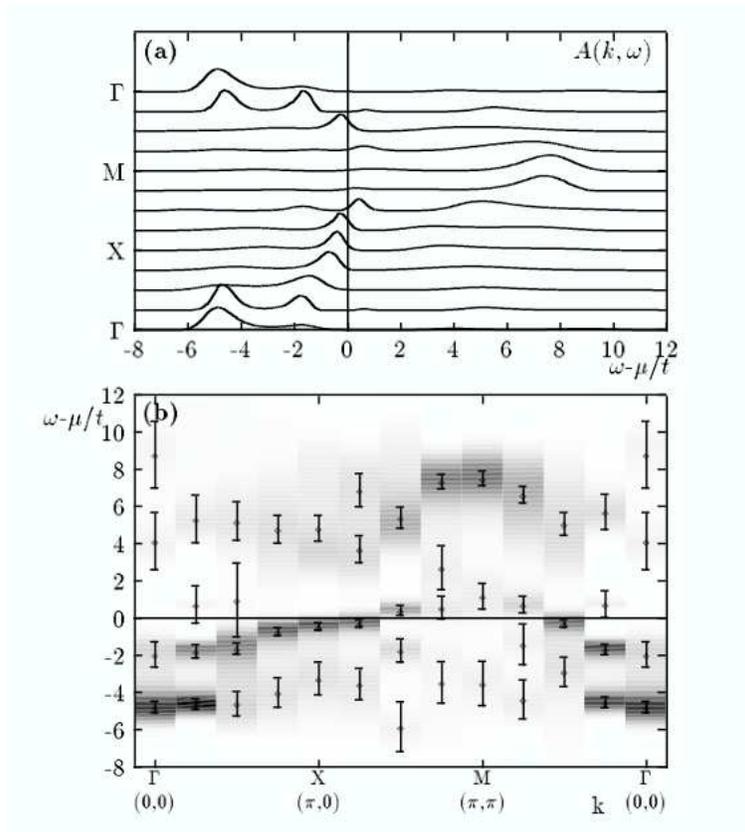}
\caption{The single-particle spectral weight $A(k,\omega)$ for the hole doped
$\langle n\rangle=0.95$ system at a temperature $T=0.33t$.  This plot is similar to
Fig.~\protect\ref{fig:15} and shows what happens as holes are doped into the Mott-Hubbard
insulator. (Preuss \etal\protect\cite{ref:A4})}
\label{fig:17}
\end{center}
\end{figure}

The left hand part of Fig.~\ref{fig:16} shows the single particle density of states for the half-filled
case.  Here, when the temperature is small compared to the exchange energy $J\sim4t^2/U$,
one clearly sees the broad upper and lower Hubbard bands and the narrow inner
bands.  When the system is hole-doped, the chemical potential moves down into the narrow coherent band
that lays below $\omega=0$ for the half-filled case and at the same time the upper coherent band
loses weight and disappears as shown on the right hand side of Fig.~\ref{fig:16}.  This is also seen in
Fig.~\ref{fig:17} which shows $A(k, \omega)$ for $\langle n\rangle=.95$
from Ref.~\cite{ref:A4}.  Here, one sees that the dispersing band below $\omega=0$ in the
insulator and the band that the holes are doped into as the system becomes metallic are quite
similar. At the same time, the narrow dispersing band that lays
just above $\omega=0$ in the insulating state has lost most of its spectral weight.

For the doped system, the fermion sign problem limited the temperature to $T=t/3$ for the
determinantal data shown in Fig.~\ref{fig:17}, but later similar determinantal Quantum
Monte Carlo runs at $T=0.25t$ and a filling of 0.95 found evidence for the formation
of a pseudogap near $(\pi,0)$.\cite{ref:A5}  In this work, the spin susceptibility was
shown to have a large spectral weight at well defined spin excitations for the doping and
temperature range in which the pseudogap appeared.  There was no pseudogap in the overdoped
$\langle n\rangle\ltwid0.8$ regions where the spectral weight of the spin susceptibility became broad and featureless.

\begin{figure}[htb]
\begin{center}
\includegraphics[width=10cm,angle=0]{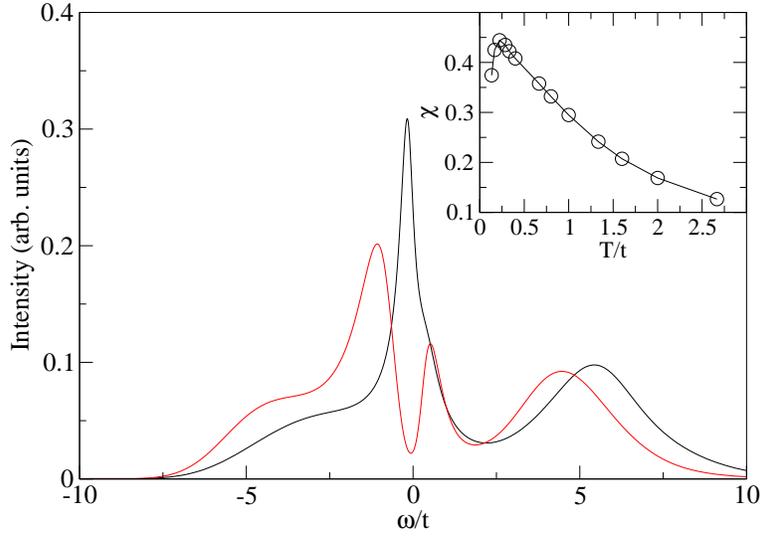}
\caption{The single particle spectral weight $A(k,\omega)$ versus $\omega$ for the antinodal
$k=(\pi,0)$ (red curve) and nodal $k=(\pi/2,\pi/2)$ (black curve) momenta of an underdoped
$\langle n\rangle=0.95$, $U=6t$ Hubbard model at $T=0.11t$.  The inset shows the temperature
dependence of the magnetic susceptibility. (M.~Jarrell)}
\label{fig:18}
\end{center}
\end{figure}

Dynamic cluster Monte Carlo calculations\cite{ref:37} with $U=6t$ and $\langle n\rangle = 0.95$ find that the
magnetic spin susceptibility exhibits a clear decrease below a temperature $T\cong 0.1t$,
as shown in the inset of Fig.~\ref{fig:18} and simulations at $T=.06t$ gave the results for $A(k,
\omega)$ shown in Fig.~\ref{fig:18}. Here, a pseudogap has opened for $k=(\pi, 0)$, while the
nodal region with $k=(\frac{\pi}{2}, \frac{\pi}{2})$ remains gapless.  In addition, a variety
of other cluster calculations\cite{ref:7,ref:15,ref:17,ref:B2} have found pseudogap behavior in both hole- and
electron-doped Hubbard models and studied its dependence on the next near-neighbor hopping
$t^\prime$.  The $t^\prime$ dependence as well as the doping dependence are consistent with
renormalization-group calculations which show the importance of umklapp scattering processes\cite{ref:34}
and the short range antiferromagnetic spin correlations.

In the next section, we turn to a discussion of the effective pairing interaction.
Specifically, the structure of the two-particle irreducible vertex and its associated
d-wave eigenfunction are analyzed.
%  We view this as central to understanding the pairing mechanism.  
%However, the reader may want to keep in mind that in this section we have
%discussed Monte Carlo and dynamic cluster calculations that found d-wave correlations
%and superconductivity for moderate values of $U$ and a DMRG calculation that at stronger
%coupling found stripes.  Based upon these results one could imagine a scenario in which
%one has d-wave pairing in the presence of dynamically fluctuating stripes.\cite{ref:B4}  Elsewhere
%in this volume the reader will find a chapter which discusses a dynamic inhomogeneity-induced
%pairing mechanism.\cite{ref:B5}

\section*{4. The Structure of the Effective Pairing Interaction}

\begin{figure}[htb]
\begin{center}
\includegraphics[width=14cm,angle=0]{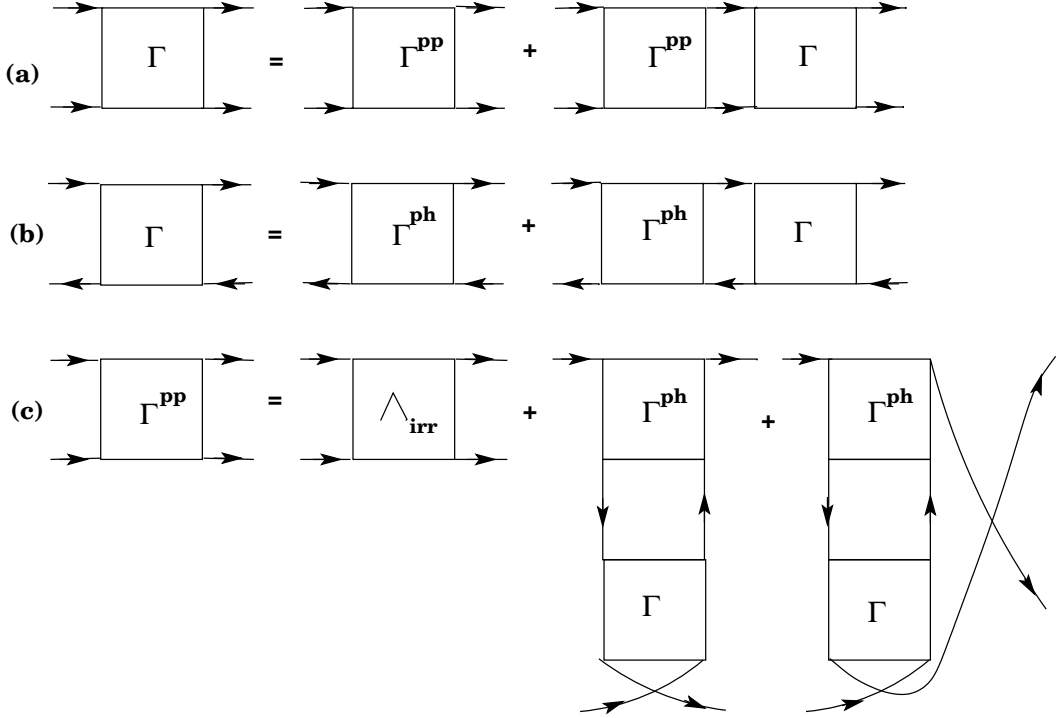}
\caption{Bethe-Salpeter equations for (a) the particle-particle and (b) the particle-hole
channels showing the relationship between the full vertex, the particle-particle irreducible
vertex $\Gamma^{pp}$, and the particle-hole irreducible vertex $\Gamma^{ph}$, respectively.
(c) Decomposition of the irreducible particle-particle vertex $\Gamma^{pp}$ into a fully
irreducible two-fermion vertex $\wedge_{\rm irr}$ plus contributions from the particle-hole
channels (Maier \etal \protect\cite{ref:21}).}
\label{fig:19}
\end{center}
\end{figure}

As discussed in Section~3, determinantal quantum Monte Carlo studies of the doped two-dimensional Hubbard model
find that $d_{x^2-y^2}$ pairing correlations develop as the temperature is lowered
and a dynamic cluster quantum Monte Carlo calculation on Betts' clusters finds evidence
for a finite temperature d-wave superconducting phase.  Here we discuss how
one can use numerical techniques to determine the structure of the interaction responsible
for the pairing.  The basic idea is to numerically calculate the four-point vertex $\Gamma$
and the single particle propagator $G$ (solid lines) shown in Fig.~\ref{fig:19}.
Then, using
the particle-particle Bethe-Salpeter equation (Fig.~\ref{fig:19}a), one can extract the two-particle
irreducible vertex $\Gamma^{\rm pp}$ which is the pairing interaction.  As we will
discuss, the four-point vertex $\Gamma$ also contains information on the particle-hole
magnetic $(S=1)$ and charge $(S=0)$ channels.  Thus, it provides a natural framework for
understanding the relationship of the pairing interaction to these other channels.

Using quantum Monte Carlo simulations, one can calculate both the one- and two-fermion
Green's functions
\begin{equation}
G(x_2, x_1)= - \langle Tc_\sigma(x_2)c^\dagger_\sigma (x_1)\rangle
\label{forty-two}
\end{equation}
and
\begin{equation}
G_2(x_4, x_3, x_2, x_1) = - \left\langle T\, c_{\sigma_4}(x_4)\, c_{\sigma_3}(x_3)
\, c^\dagger_{\sigma_2}(x_2)\, c^\dagger_{\sigma_1}(x_1)\right\rangle
\label{forty-three}
\end{equation}
Here, $c^\dagger_\sigma(x_\ell)$ creates an electron with spin $\sigma$ at site $x_\ell$
and imaginary time $\tau_\ell$.  $T$ is the usual $\tau$-ordering operator and we have
suppressed the $\sigma$ indices. Fourier transforming on both the space
and imaginary time variables, one obtains $G(p)$ and
\begin{eqnarray}
G_2(p_4, p_3, p_2, p_1) & = & - G(p_1)G(p_2)(\delta_{p_1, p_4} \delta_{p_2, p_3}-
 \delta_{p_1, p_3} \delta_{p_2, p_4})
\nonumber\\
& + & \frac{T}{N}\ \delta_{p_1 + p_2, p_3+p_4} G(p_4)\, G(p_3) \Gamma (p_4,
p_3;\ p_2, p_1)\, G(p_2)\, G(p_1)
\label{forty-four}
\end{eqnarray}
with $p=({\bf p}, i\omega_n)$.  Then, using the Monte Carlo results for $G$ and $G_2$,
one can determine the four-point vertex $\Gamma$ from Eq.~(\ref{forty-four}).

Given $\Gamma$ and $G$, one can solve the Bethe-Salpeter equations shown in Fig.~\ref{fig:19}a and b to
obtain the irreducible particle-particle and particle-hole vertices $\Gamma^{\rm pp}$ and
$\Gamma^{\rm ph}$.  For example, in the zero center of mass and energy channel, the
particle-particle Bethe-Salpeter equation shown in Fig.~\ref{fig:19}a gives
\begin{equation}
\Gamma(p^\prime | p) = \Gamma^{\rm pp} (p^\prime | p) - \frac{T}{N}
\sum_k \Gamma^{\rm pp} (p^\prime | k)\, G_\uparrow (k)\, G_\downarrow (-k)
\, \Gamma (k|p)
\label{forty-five}
\end{equation}
with $\Gamma(p^\prime|p)=\Gamma(p^\prime, -p^\prime;\ p, -p)$.  Given
$\Gamma$ and $G$, one can then determine the irreducible particle-particle vertex
$\Gamma^{\rm pp}$.  This procedure is essentially the opposite of what one does in the
traditional diagrammatic approach.  There, one introduces an approximation for the
irreducible vertex $\Gamma^{\rm pp}$ and solves Eq.~(\ref{forty-five}) for $\Gamma$.  Here, we use Monte Carlo results
for $\Gamma$ and $G$ and solve Eq.~(\ref{forty-five}) for $\Gamma^{\rm pp}$. The Monte Carlo results
for $\Gamma$ satisfy crossing symmetry and $\Gamma^{\rm pp} (p^\prime |p)$ obtained from
Eq.~(\ref{forty-five}) is the effective particle-particle interaction.  There is no
approximation except for the fact that a finite lattice is used and one has
the usual statistical Monte Carlo errors (and the small systematic finite $\Delta\tau$
errors which can be eliminated by extrapolating $\Delta\tau\to0$).

\begin{figure}[htb]
\begin{minipage}[b]{8cm}
%\begin{center}
\centerline{\includegraphics[width=7.5cm]{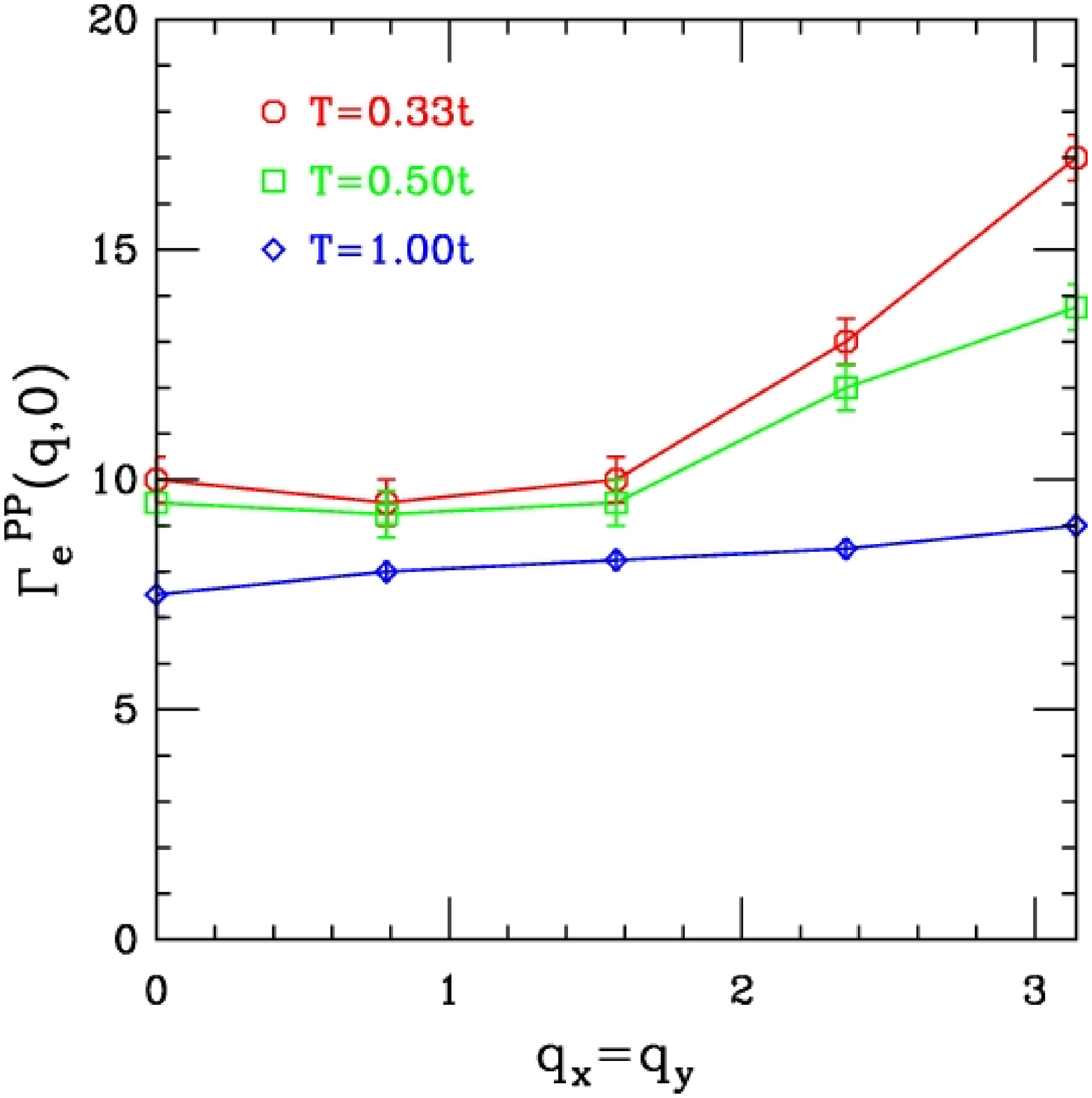}\phantom{xxx}}
\caption{The even irreducible particle-particle vertex $\Gamma^{pp}_e({\bf q},\omega_m=0)$ for
${\bf q}={\bf p}^\prime-{\bf p}$ and ${\bf p}=(\pi,0)$ versus
momentum transfer ${\bf q}$ along the $(1,1)$ direction.  Here $U=4t$ and $\langle n\rangle=0.875$.
As the temperature decreases below the temperature where spin-spin correlations develop, the
strength of the interaction is enhanced at large momentum transfers. (Bulut \etal\protect\cite{ref:20})}
\label{fig:20}
%\end{center}
\end{minipage}
\hfill
\begin{minipage}[b]{8cm}
%\begin{center}
\centerline{\includegraphics[width=7.5cm]{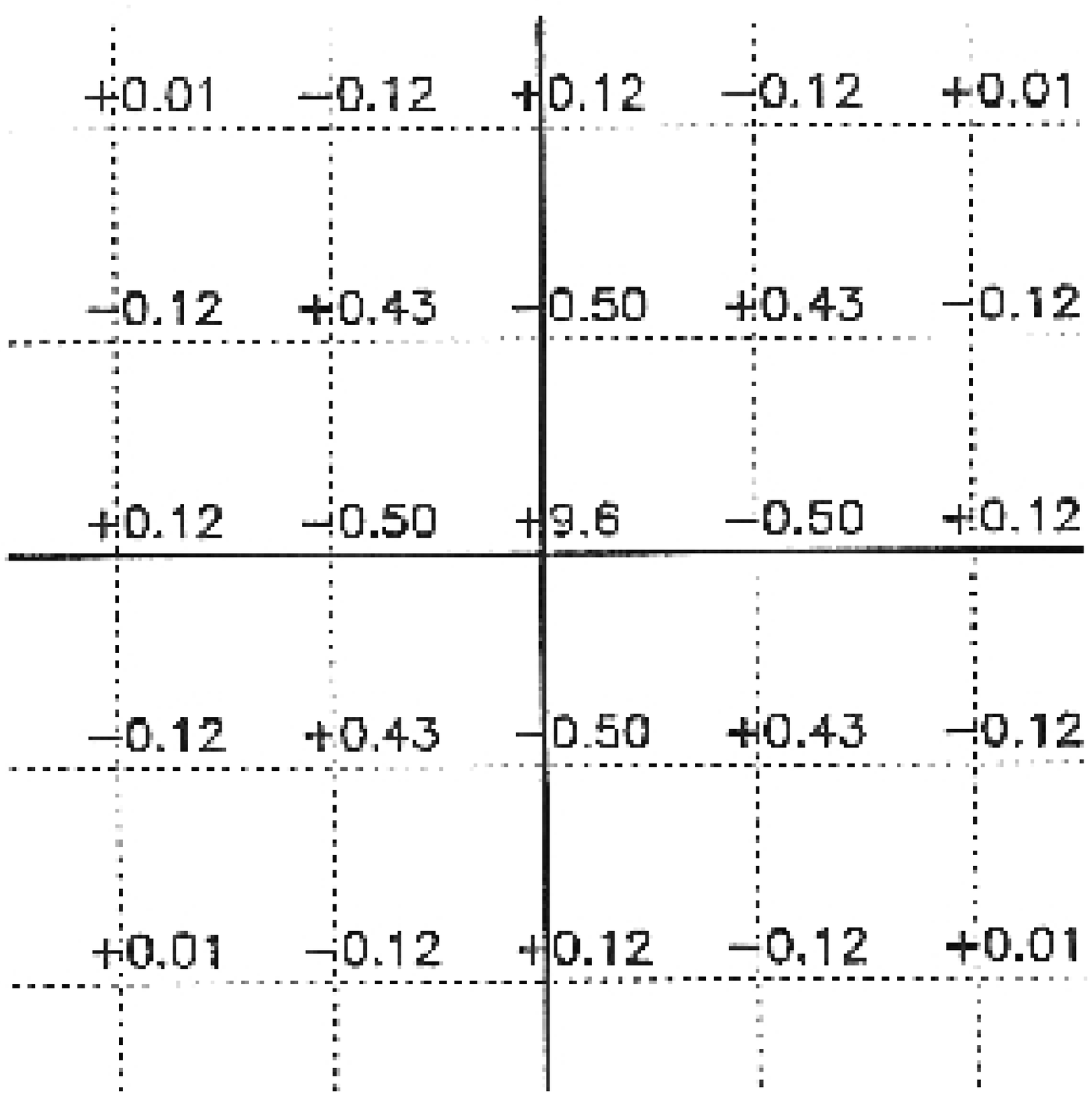}}
\vspace{0.75cm}
\caption{The real-space structure of $\Gamma^{pp}_e({\bf R})$ at a temperature $T=0.25t$ for
$U=4t$ and $\langle n\rangle=0.87$.  When the singlet electron pair is separated by one
lattice spacing, ${\bf R=x}$ or {\bf y}, the interaction is attractive, while it is strongly repulsive when ${\bf R}=0$
and the pair occupy the same site. (Bulut \etal\protect\cite{ref:20})}
\label{fig:21}
%\end{center}
\vspace{0.5cm}
\end{minipage}
\end{figure}

The dominant pairing response, at low temperatures, is found to occur in the even frequency
$d_{x^2-y^2}$ channel.  Since this channel is even in both the relative frequency and momentum,
it must be a spin singlet.  Note that there are also spin singlet pairing channels which are
odd in the relative frequency and momentum.  However, the pairing instability in the doped
Hubbard model comes from the even frequency and even momentum part of the irreducible
particle-particle vertex.
\begin{equation}
\Gamma_e^{\rm pp} (p^\prime |p) = \frac{1}{2}\ \left[\Gamma^{\rm pp}(p^\prime|p) +
\Gamma^{\rm pp} (-p^\prime|p)\right]\, .
\label{forty-six}
\end{equation}
Determinantal quantum Monte Carlo results\cite{ref:20} for $\Gamma^{\rm pp}_e (p^\prime |p)$ obtained
from an 8$\times$8 lattice with $U=4t$ and $\langle n\rangle=0.87$ are shown in Fig.~\ref{fig:20}.
Here, $\Gamma^{\rm pp}_e (p^\prime |p)$ is plotted for various temperatures as a function
of ${\bf q}={\bf p}^\prime-{\bf p}$ with ${\bf p}=(\pi,0)$ and $\omega_n=\omega_{n^\prime}=\pi T$. One sees that as the
temperature is lowered, $\Gamma^{\rm pp}_e$ peaks at large momentum transfers.
The size of the effective pairing interaction $\Gamma^{\rm pp}_e$ also depends
upon the energy transfer $\omega_m=\omega_{n^\prime}-\omega_n$, and falls off with
$\omega_m$ on a scale set by the characteristic spin-fluctuation energy.

To obtain a more intuitive picture of the way in which the local repulsive
$Un_{i\uparrow} n_{i\downarrow}$ Hubbard interaction can lead to an effective attractive
pairing interaction in the singlet channel, it is useful to construct the real space
Fourier transform
\begin{equation}
\Gamma^{\rm pp}_e ({\bf R}) = \frac{1}{N^2} \sum_{{\bf p},
{\bf p}^\prime} e^{i({\bf p}^\prime-{\bf p})\cdot {\bf R}}
\ \Gamma^{\rm pp}_e ({\bf p}^\prime, i\pi T; {\bf p}, i\pi T)\, .
\label{forty-seven}
\end{equation}
Values for $\Gamma^{\rm pp}_e ({\bf R})$ are shown in Fig.~\ref{fig:21}, with the distance ${\bf R}$ between the two fermions
measured from the central point.  If two fermions occupy the same site, spin-up and
spin-down, $\Gamma^{\rm pp}_e ({\bf R}=0) \simeq 9.6t$. That is, the effective pairing interaction
is even more repulsive than the bare $U=4t$ onsite Coulomb interaction. However, if two
fermions in a singlet state are on near-neighbor sites, the effective interaction
$\Gamma^{\rm pp}_e ({\bf R}=\hat {\bf x}\ {\rm or}\ \hat {\bf y}) \simeq -0.5t$ is attractive.
%In a rough
%way, the spatial structure of $\Gamma^{\rm pp}_s({\bf R})$ can be viewed as a large Friedel
%oscillation produced around a dynamic hole by the underlying onsite Hubbard $U$. However, it
%is important to keep in mind that the interaction is dynamic with a time scale related to the
%spin fluctuation response.

\begin{figure}[htb]
\begin{minipage}[b]{8cm}
%\begin{center}
\centerline{\includegraphics[width=7.5cm]{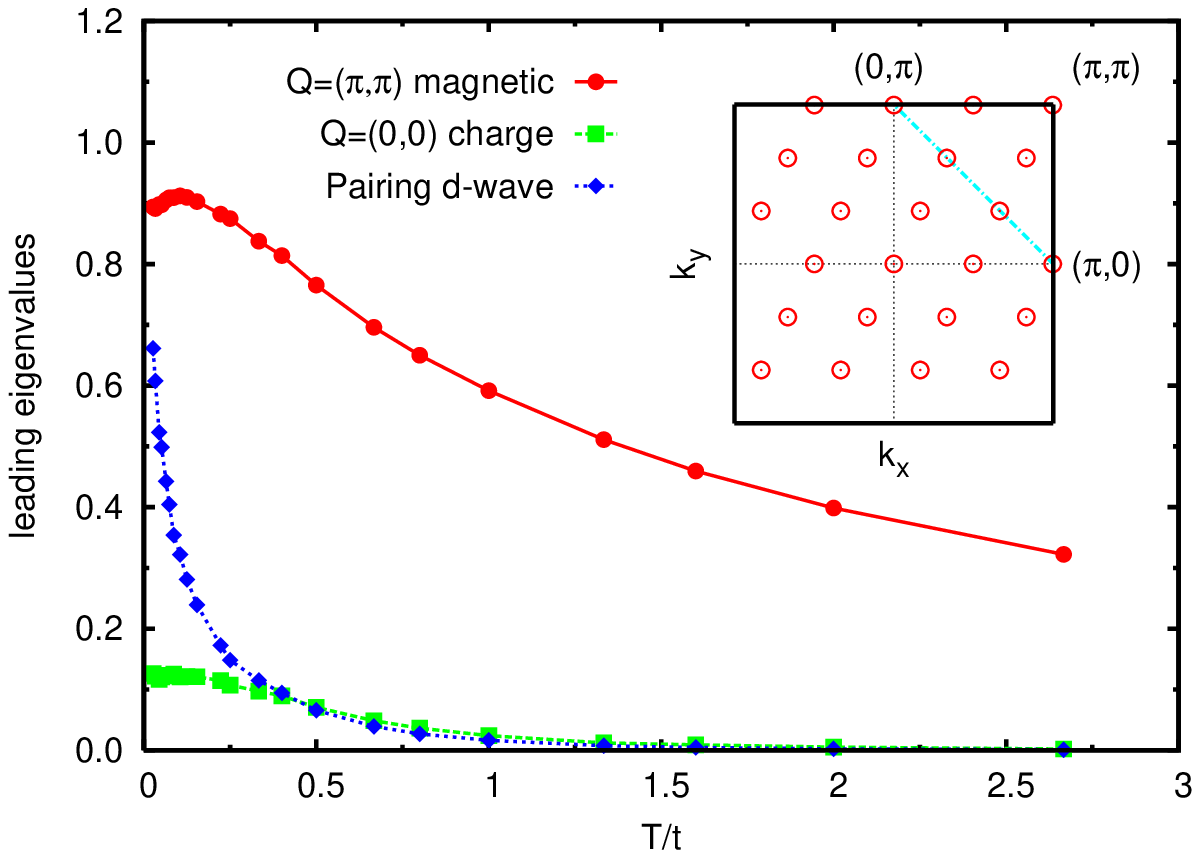}\phantom{xx}}
%\vspace{0.1cm}
\caption{Leading eigenvalues of the Bethe-Salpeter equations
in various channels for $U/t=4$ and a site occupation $\langle n\rangle=0.85$.  The
${\bf Q}=(\pi,\pi)$, $\omega_m=0$, $S=1$ magnetic eigenvalue is seen to peak at
low temperatures.  The leading eigenvalue in the even singlet ${\bf Q}=(0,0)$, $\omega_m=0$
particle-particle channel has $d_{x^2-y^2}$ symmetry and increases toward 1 at low
temperatures.  The largest charge density eigenvalue occurs in the
${\bf Q}=(0,0)$, $\omega_m=0$ channel and saturates at a small value.  The inset shows the
distribution of $k$-points for the 24-site cluster. (Maier \etal\protect\cite{ref:21})}
\label{fig:22}
%\end{center}
\vspace{1.6cm}
\end{minipage}
\hfill
\begin{minipage}[b]{8cm}
%\begin{center}
\centerline{\includegraphics[width=7.8cm]{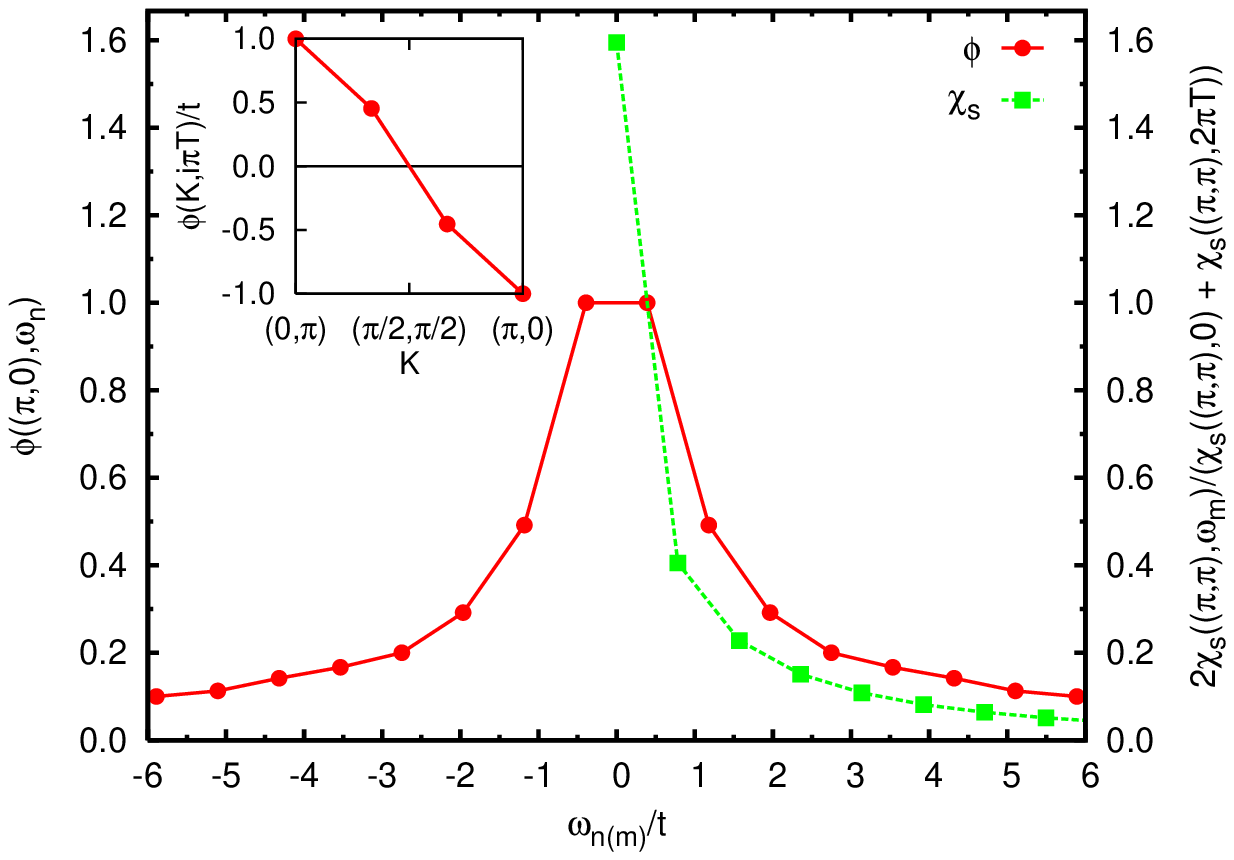}}
\caption{The Matsubara frequency dependence of the eigenfunction
$\phi_{d_{x^2-y^2}}({\bf K},\omega_n)$ of the leading particle-particle
eigenvalue of Fig.~\protect\ref{fig:22} for ${\bf K}=(\pi,0)$ normalized to
$\phi({\bf K},\pi T)$ (red).  Here, $\omega_n=(2n+1)\pi T$ with $T=0.125t$.  For
comparison, the Matsubara frequency dependence of the normalized magnetic spin susceptibility
$2\chi({\bf Q},\omega_m)/[\chi({\bf Q},0)+\chi({\bf Q},2\pi T)]$ for ${\bf Q}=(\pi,\pi)$
versus $\omega_m=2m\pi T$ is also shown (green).  In the inset,
the momentum dependence of the eigenfunction $\phi_{d_{x^2-y^2}}({\bf K},\pi T)$
normalized to $\phi_{d_{x^2-y^2}}((0,\pi),\pi T)$ shows its $d_{x^2-y^2}$ symmetry.  Here,
$\omega_n=\pi T$ and the momentum values correspond to values of ${\bf K}$ which lay along
the dashed line shown in the inset of Fig.~\protect\ref{fig:22}. (Maier \etal\protect\cite{ref:21})}
\label{fig:23}
%\end{center}
\end{minipage}
\end{figure}

In order to determine the structure of the pairing correlations which are produced by
$\Gamma^{\rm PP}_e$, we turn to the homogenous Bethe-Salpeter equation
\begin{equation}
-\frac{T}{N} \sum_{p^\prime} \Gamma_e^{\rm pp} (p|p^\prime)\, G_\uparrow (p^\prime)
\, G_\downarrow (-p^\prime)\, \phi_\alpha (p^\prime) = \lambda_\alpha \phi_\alpha(p)\, .
\label{forty-eight}
\end{equation}
The temperature dependence of the leading eigenvalue in the particle-particle channel is
plotted versus the temperature in Fig.~\ref{fig:22}. When this eigenvalue reaches 1, it
signals an instability into a superconducting phase.  Here, $U=4t$ with $\langle
n\rangle=0.85$ and we are showing results obtained using the dynamic cluster approximation\cite{ref:21}
for the 24-site $k$-cluster discussed in Section 3. The distribution of ${\bf k}$ points for
the 24-site cluster is shown in the inset of Fig.~\ref{fig:22}. Similar results for $T\geq 0.25t$ have
been obtained using the determinantal Monte Carlo algorithm on an 8$\times$8 lattice.\cite{ref:20}

The eigenfunction corresponding to the leading particle-particle eigenvalue is a singlet and
its $K$ dependence, plotted in the inset of Fig.~\ref{fig:23}, shows that it has $d_{x^2-y^2}$ symmetry. The frequency
dependence of this eigenfunction at the antinodal point $K=(\pi, 0)$ is shown in 
the main part of Fig.~\ref{fig:23}.  Here,
$\phi((\pi,0),\omega_n)$ has been normalized so that at $\omega_n=\pi T$ its value is 1.
It is even in $\omega_n$ as it must be for a d-wave singlet to satisfy the Pauli principle.
Also shown in this figure is the $\omega_m$-dependence of the $Q=(\pi, \pi)$ spin
susceptibility $\chi(Q, \omega_m)$ normalized by $(\chi(Q,0)+\chi(Q,2\pi T))/2$ for comparison
with $\phi((\pi,0),\omega_n)$. The boson Matsubara frequency dependence, $\omega_m=2m\pi
T$, of the susceptibility is seen to interlace with the fermion, $\omega_n=(2n+1)\pi T$,
dependence of the eigenfunction.  The momentum and frequency dependence of
$\phi_{d_{x^2-y^2}} (K, \omega)$ reflects the structure of the pairing interaction
$\Gamma^{\rm pp}_e$. The numerical results show that $\Gamma^{\rm pp}_e$ is an increasing function of momentum
transfer and is characterized by a similar energy scale to that which enters the spin susceptibility
$\chi(Q,\omega_m)$.

In a similar way, one can use $\Gamma$ and $G$ to solve for the irreducible particle-hole
vertex $\Gamma^{\rm ph}$ shown in Fig.~\ref{fig:19}b. The homogenous Bethe-Salpeter equation for the channel
with center-of-mass momentum $Q$, Matsubara frequency $\omega_m=0$ and $z$-component of spin $S_z=0$ is
\begin{equation}
-\frac{T}{N} \sum_{k^\prime} \Gamma^{\rm ph} (k+Q, k; k^\prime+Q, k^\prime)\, G_\uparrow
(k^\prime + q)\, G_\downarrow (k^\prime)\, \phi_{Q\alpha} (k^\prime)=\lambda_\alpha (Q)\,
\phi_{Q\alpha} (k)\, .
\label{forty-nine}
\end{equation}
The leading eigenvalue in the particle-hole channel occurs for $Q=(\pi, \pi)$ for the
24-site
$k$-cluster and carries spin 1.  Earlier determinantal quantum Monte Carlo studies\cite{ref:A2} on
8$\times$8 lattices show that for this doping the peak response is, in fact, slightly shifted
from $(\pi, \pi)$, but the 24-site $k$-cluster used in the dynamic cluster calculation lacks
the resolution to show this.  As seen in Fig.~\ref{fig:22}, for this doping, the antiferromagnetic eigenvalue initially
grows as the temperature is reduced, peaking at low temperatures.  The largest eigenvalue in
the $S=0$ charge density channel occurs for $Q=(0,0)$ and $\omega_m=0$.  Its temperature dependence
is also plotted in Fig.~\ref{fig:22}.

\begin{figure}[htb]
\centerline{\includegraphics[width=12cm]{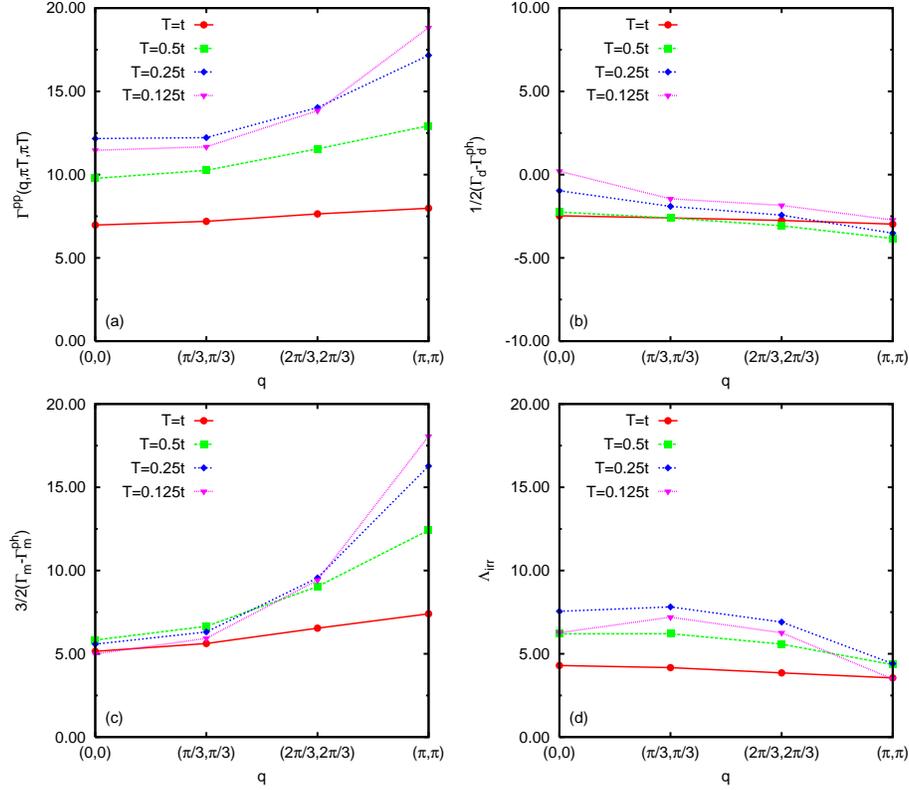}}
\caption{(a) The irreducible particle-particle vertex $\Gamma^{pp}_e$ versus ${\bf q}=
{\bf K}-{\bf K}^\prime$ for various temperatures with $\omega_n=\omega_{n^\prime}=\pi T$.
Here, ${\bf K}=(\pi,0)$ and ${\bf K}^\prime$ moves along the momentum values of the
24-site cluster which lay on the dashed line shown in the inset of Fig.~\ref{fig:22}.  Note
that the interaction increases with the momentum transfer as expected for a d-wave pairing
interaction.  (b) The {\bf q}-dependence of the fully irreducible two-fermion vertex
$\wedge_{\rm irr}$. (c) The {\bf q}-dependence of the charge density $(S=0)$ channel
$\frac{1}{2}\Phi_d$ for the same set of temperatures. (d) The {\bf q}-dependence of the
magnetic $(S=1)$ channel $\frac{3}{2}\Phi_m$. (Maier \etal\protect\cite{ref:21})}
\label{fig:24}
\end{figure}

Returning to the question of the structure of the irreducible particle-particle vertex
$\Gamma^{\rm pp}_e$, we have seen that $\Gamma^{\rm pp}_e$ peaks at large momentum
transfers and has a frequency dependence reflected in $\Phi_{d_{x^2-y^2}}(K,\omega_n)$
which is similar to the spin susceptibility.  However, we would like to understand one
further aspect. Is the dominant contribution to the $d_{x^2-y^2}$ pairing interaction
associated with an $S=1$ particle-hole channel? Alternatively, for example, one could
have a charge density $S=0$ channel or a more complicated multiparticle-hole exchange
process such as that suggested by the spin-bag picture.\cite{ref:KS}

In order to address this, we will make use of the representation of $\Gamma^{\rm pp}$
shown diagrammatically in Fig.~\ref{fig:19}c.  Here, $\Gamma^{\rm pp}$ is decomposed into a fully
irreducible vertex $\wedge_{\rm irr}$ plus contribution from particle-hole exchange channels.
Because of the spin rotation invariance of the Hubbard model, one can separate the
particle-hole channels into a charge density $S=0$ contribution
and a spin $S=1$ magnetic part.  For the even frequency and even momentum (singlet pairing)
part of the irreducible particle-particle vertex, Eq.~(\ref{forty-six}), one has
\begin{equation}
\Gamma^{\rm pp}_e (p^\prime|p) = \Lambda_{\rm irr} (p^\prime |p) + \frac{1}{2}
\ \Phi_d (p^\prime,p) + \frac{3}{2}\ \Phi_m(p^\prime,p)\, .
\label{fifty}
\end{equation}
The subscripts $d$ and $m$ denote the charge density $(S=0)$ and magnetic $(S=1)$
particle-hole channels respectively, with
\begin{eqnarray}
\Phi_{d/m}(p^\prime,p) & = & \frac{1}{2}\biggl[\Gamma_{d/m}(p^\prime-p;p,-p^\prime)-
                             \Gamma^{ph}_{d/m}(p^\prime-p;p,-p^\prime) \nonumber \\
                         & + & \Gamma_{d/m}(p^\prime+p;-p,-p^\prime)-
                                                   \Gamma^{ph}_{d/m}(p^\prime+p;-p,-p^\prime)\biggr]
\label{fifty-one}
\end{eqnarray}
Here, on the right hand side, the center of mass and relative wave vectors and frequencies
in these channels are labeled by the first, second and third arguments, respectively.

Results for the irreducible particle-particle interaction $\Gamma^{\rm pp}_e$ obtained from
the 24-site dynamic cluster approximation are shown in Fig.~\ref{fig:24}. As we have seen, when the
temperature is lowered, $\Gamma^{\rm pp}_e$ increases as the momentum transfer
${\bf q=p^\prime-p}$ increases.  Using the results for $\Gamma^{\rm ph}$, $\Gamma$, and $G$ one
can calculate the contributions $\Phi_d$ from the $S=0$ charge density and $\Phi_m$ for the $S=1$ magnetic
channels. Subtracting these
from $\Gamma^{\rm pp}_e$ gives $\Lambda_{\rm irr}$ and results for each of these contributions
are shown in Fig.~\ref{fig:24}. The dominant $d_{x^2-y^2}$ pairing contribution to $\Gamma^{\rm pp}_e$
clearly comes from the $S=1$ channel.

\begin{figure}[htb]
\begin{minipage}[b]{8cm}
%\begin{center}
\centerline{\includegraphics[width=8.0cm]{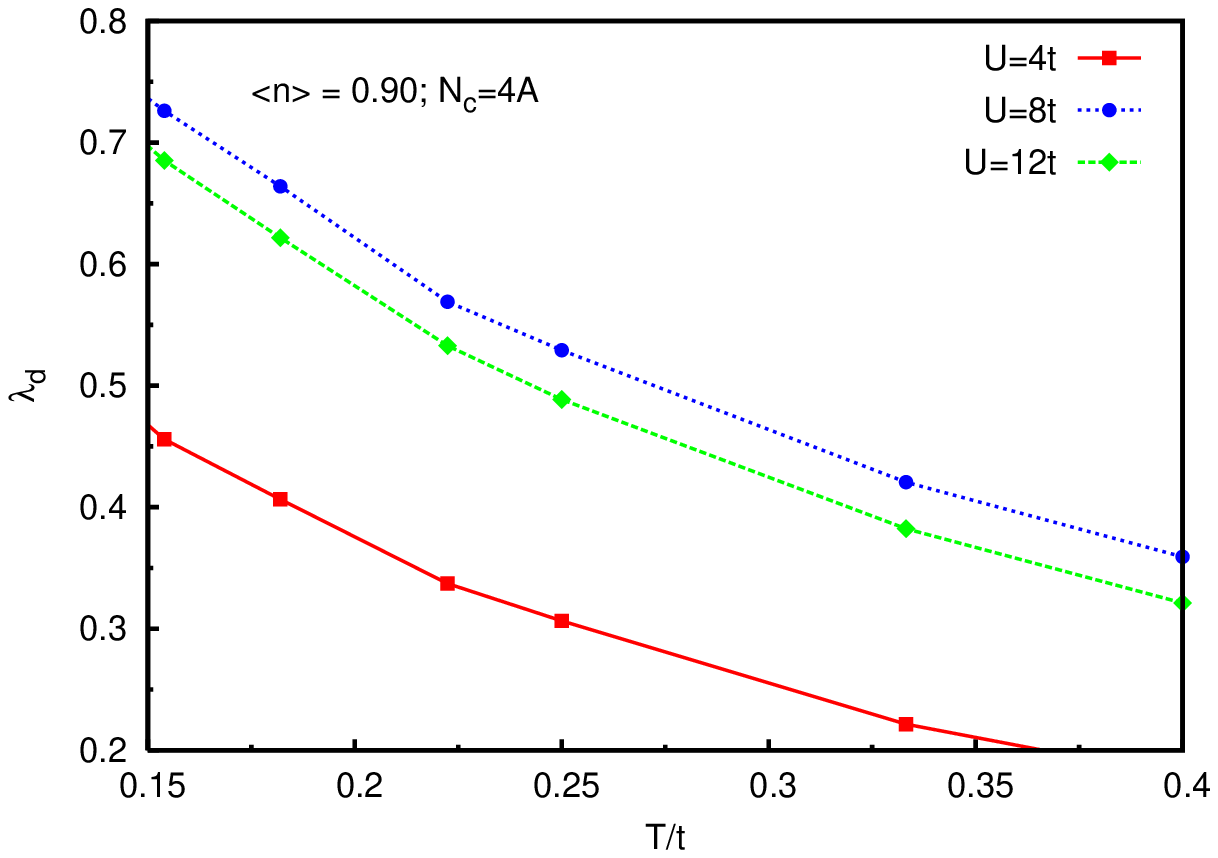}\phantom{xxx}}
\caption{The $d_{x^2-y^2}$-wave eigenvalue $\lambda_d$ versus temperature $T/t$
for $\langle n\rangle=0.85$ with $U=4t$ (red), $U=8t$ (blue) and $U=12t$ (green).
These results were obtained for a 4-site $k$-cluster (Maier \etal\protect\cite{ref:MJS}).}
\label{fig:25}
%\end{center}
\vspace{1.4cm}
\end{minipage}
\hfill
\begin{minipage}[b]{8cm}
\begin{center}
\centerline{\includegraphics[width=8.0cm]{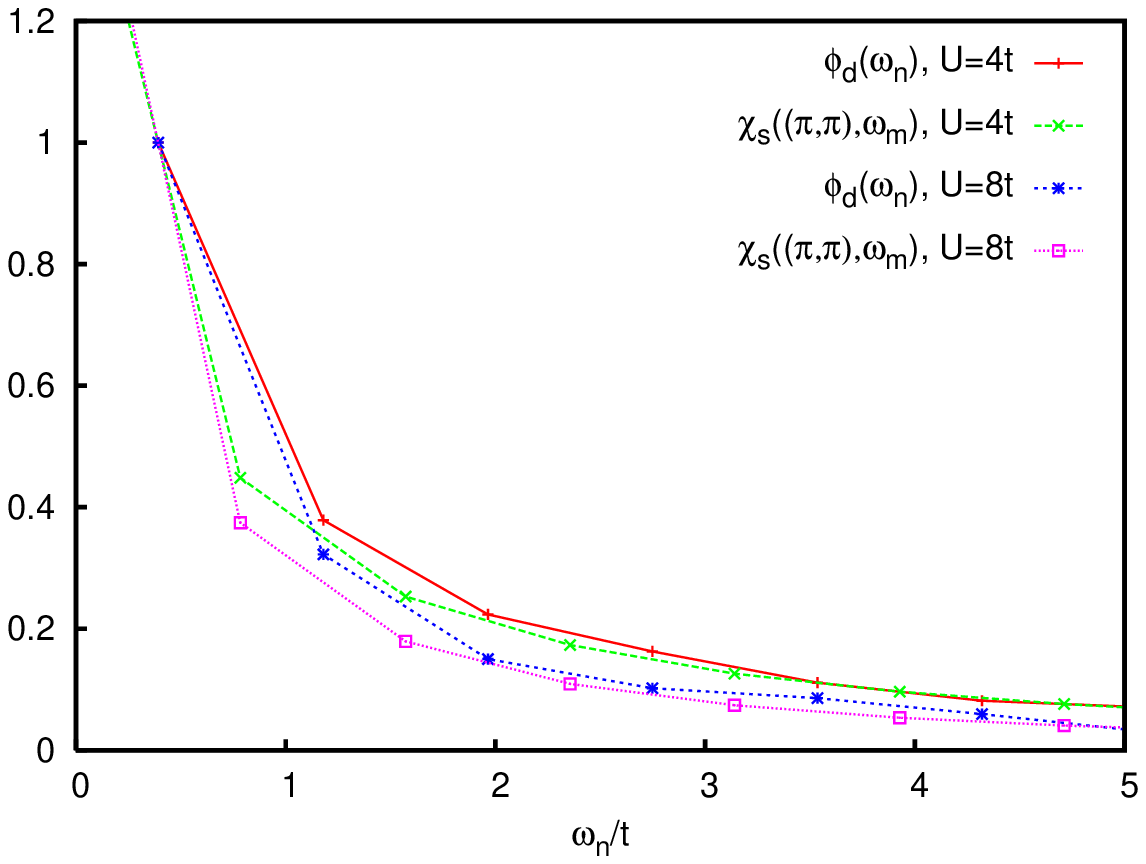}\phantom{xx}}
\caption{The Matsubara frequency dependence of the $d_{x^2-y^2}$ eigenfunction $\phi_d(K,\omega_n)$
with $K=(\pi,0)$ for $U=4t$, $8t$ and $12t$.
The spin susceptibility $\chi(Q,\omega_m)$ for $Q=(\pi,\pi)$ is also shown for comparison.
Here, $\phi_d$ and $\chi$ are normalized in the same
way as the results shown in Fig.~\protect\ref{fig:24} (Maier \etal\protect\cite{ref:MJS}).}
\label{fig:26}
\end{center}
\end{minipage}
\end{figure}

At larger values of $U$, 4-site $k$-cluster calculations\cite{ref:MJS} of the temperature
dependence of the $d_{x^2-y^2}$ eigenvalue for $\langle n\rangle=0.85$ and $U=4t$, $8t$
and $12t$ are shown in Fig.~\ref{fig:25}.  Over the temperature range\cite{ref:Ref10} shown
in Fig.~\ref{fig:25}, the $d_{x^2-y^2}$ eigenvalue is largest for $U=8t$.  This is consistent
with the 2-leg ladder result shown in Fig.~\ref{fig:12a} and the expectation that the maximum
transition temperature occurs for $U$ of order the bandwidth.
The $d_{x^2-y^2}$ eigenfunction
$\phi_{d_{x^2-y^2}}(K,\omega_n)$ has the expected d-wave $K$ dependence and its Matsubara
frequency dependence for $U=4t$ and $8t$ are shown in Fig.~\ref{fig:26}.  Here, as before, we also show the
$\omega_m$ dependence of the spin susceptibility $\chi(Q,\omega_m)$.  As $U$ increases,
both $\phi_{d_{x^2-y^2}}(K,\omega_n)$ and $\chi(Q,\omega_m)$ fall off more rapidly,
reflecting the reduction in the frequency scale set by $J\sim4t^2/U$. 

\section*{5. Conclusions}

The numerical studies of the Hubbard model that we have reviewed show that it exhibits
the basic properties that are observed in the cuprate materials: antiferromagnetism,
$d_{x^2-y^2}$-pairing, stripes and pseudogap phenomena.  Numerical methods have also been
used to study the structure of the interaction responsible for pairing in the Hubbard model.
As discussed in Section~4, this can be done by directly calculating the irreducible
particle-particle vertex $\Gamma^{pp}$ or by studying the momentum and frequency dependence
of the gap function $\phi_{d_{x^2-y^2}}(K,\omega)$.  The decomposition of $\Gamma^{pp}$ showed that the
dominant pairing interaction arose from a spin-one particle-hole exchange.  The strength
of $\Gamma^{pp}$ was found to increase with momentum transfer leading to $d_{x^2-y^2}$-pairing.
Alternately, the $(\cos K_x-\cos K_y)$ momentum dependence of the gap function $\phi_{d_{x^2-y^2}}(K,\omega)$
and the similarity of its $\omega_n$ dependence to that of the $Q=(\pi,\pi)$ spin susceptibility
leads to the same conclusion: the pairing interaction in the doped Hubbard model is repulsive on
site, attractive between near-neighbor sites and retarded on a time scale set by the inverse
of the spin-fluctuation spectrum.  It is important to recognize that this spectrum includes a
particle-hole continuum.

Now, one can ask whether this interaction is actually the mechanism responsible for pairing in the high $T_c$ cuprate
materials and how one would know this from experiments?  As far as the momentum dependence of
the interaction is concerned, ARPES studies\cite{ref:36} of the $k$-dependence of the gap along with a
variety of transport\cite{ref:38} and phase dependent studies\cite{ref:39,ref:40} provide strong evidence for the nodal d-wave
character of the gap.  While it is known that the chains in YBCO lead to an admixture of
s-wave \cite{ref:kak,ref:41,ref:42} and the momentum regions probed are primarily along the fermi surface, there is good
reason to believe from the observed $k$-dependence of the gap that the pairing interaction
is indeed repulsive on site and attractive for singlets formed between near neighbor sites.
It will be interesting to compare calculations for an orthohrombic Hubbard model with experiments
\cite{ref:41,ref:42}, to see if the observed $k$-dependence of the gap can provide additional
help in identifying the pairing mechanism.

Another characteristic of the interaction is its frequency dependence.  Here, less is known
but it seems likely that the frequency dependence of the gap and renormalization parameter
will provide important insight into the mechanism.  As one knows,
it was the frequency dependence of the gap for the traditional low $T_c$ superconductors
that provided the ultimate fingerprint identifying the phonon exchange pairing interaction, although
at the time few doubted that this was the mechanism.  In the high $T_c$ case, the initial
hope was that the d-wave momentum dependence of the gap would provide a sufficiently precise
fingerprint.  However, this has not been the case.  For example, the exchange of $B_{1g}$ phonons is known
to favor d-wave pairing,\cite{ref:C2,ref:43}, although its overall contribution to $T_c$ is
small within the standard theory.  A two-band Cu-O model, in which fluctuations in circulating
currents provide a d-wave pairing mechanism, has also been proposed.\cite{ref:44}  Even within the
framework of the Hubbard model there are different views regarding the dynamics.  In the
``Plain-Vanilla-RVB" picture,\cite{ref:Anderson1} it has been suggested that the dynamics is
set by an energy scale associate with the Mott-Hubbard gap.\cite{ref:Anderson2}  However, our
numerical results support a picture in which the dominant contributions come from particle-hole
excitations within the relatively narrow band that the doped holes enter giving an energy scale
of several times $J$.  While the spectrum of these excitations extends down to zero energy, the main
strength is associated with a broad spin-fluctuation continuum.\cite{ref:Scalapino,ref:Pines}
Thus it seems likely that the dynamics will again be important in identifying the mechanism.

In addition to the traditional electron tunneling\cite{ref:45} and infrared conductivity\cite{ref:46} measurements, ARPES
experiments provide an important tool for probing the frequency dependence of the renormalization
parameter and the gap.  Advances in the energy and momentum resolution of both ARPES\cite{ref:36} and
neutron scattering\cite{ref:NS} along with material preparation techniques that allow ARPES and neutron
scattering to be done on the same material are opening new opportunities.  Various RPA-BCS
approximations have been used to model both the ARPES\cite{ref:49,ref:50} and neutron
scattering data.\cite{ref:51,ref:52}  One would clearly
like to extend the numerical Hubbard model studies so that they can be used in making
such experimental comparisons.

Finally, in addition to the frequency and momentum dependence of the interaction, there is
the question of its strength.  The estimate for the transition temperature in Sec.~3 with
$U=4t$ was relatively small.  As discussed, we believe that for larger values of $U$ (of
order the bandwidth) and a more optimal bandstructure, $T_c$ will increase.  Beyond this,
the actual Cu-O structure has additional exchange paths and it is known that $t-J-U$
Hubbard ladders can exhibit stronger pairing correlations.\cite{ref:Daul}  Nevertheless,
the question of the strength of the pairing interaction remains.  It is not that several
times $J$ isn't a wide spectral range compared to the phonon scale of the traditional low
temperature superconductors or that the system isn't strongly coupled with $U$ of order
the bandwidth.  Rather it is that the strong coupling has created a delicately balanced
system.\cite{ref:DJS}  As discussed in Sec.~3, different numerical methods on different
lattices find evidence in one case for $d$-wave pairing and in another for stripes.
Thus small changes in local parameters may alter the nature of the correlations and
there is a question regarding the role of inhomogeneity in the cuprates.  An interesting
theory of ``dynamic inhomogeneity-induced pairing" is discussed in another chapter of this
treatise.\cite{ref:B5}  In this approach, pairing from repulsive interactions appears as a
mesoscopic effect and the phenomena of high temperature superconductivity is viewed as
arising from the existence of mesoscale structures.\cite{ref:B5,ref:B4}  Recent STM
measurements of impurities and inhomogeneities in BSCCO are providing important new information
on the question of the local modulation of the pairing and its strength.\cite{ref:Yaz,ref:Hud,ref:Pan}

Thus, two decades after Bednorz' and M\"uller's\cite{ref:BM} discovery of the high $T_c$
cuprates the question of the pairing mechanism remains open.  However, it is clear that the
desire to understand these materials has driven dramatic advances in the experimental energy
and momentum resolution of ARPES and neutron scattering and the energy and spatial resolution
of STM.  It was also largely responsible for the development of a variety of numerical techniques
which are providing new insights into the electronic properties of a wide class of strongly
correlated materials.

\section*{Acknowledgement}

I would like to acknowledge past graduate students N.~Bulut, J.M.~Byers, M.~Jarrell,
E.~Loh, R.~Melko, R.M.~Noack, S.~Quinlan, and R.~Scalettar and postdocs F.~Assaad,
N.E.~Bickers, L.~Capriotti, E.~Dagotto, T.~Dahm, J.~Freericks, M.E.~Flatte, R.~Fye,
J.~Hirsch, M.~Imada, P.~Monthoux, A.~Moreo, M.~Salkola, H.B.~Schuttler and S.R.~White
who have been willing to teach me new things for so long.  I would also like to acknowledge
the pleasure I have had discussing and working on this problem with A.V.~Balatsky, W.~Hanke,
P.~Hirschfeld, S.A.~Kivelson, T.~Maier and D.~Poilblanc.  Finally I want to thank my
long time collaborator R.L.~Sugar for his insights and encouragement.  This work was
supported by NSF Grant DMR02-11166 and the Department of Energy under FG02-03ER46048.

\section*{Addendum}

This manuscript was written almost a year ago and will appear as Chapter 13 in the ``Handbook of High Temperature Superconductivity,"
edited by J.R.~Schrieffer and published by Springer.  It is narrowly focused on the results obtained from numerical studies of one
particular model, the Hubbard model.  Here in this addendum\cite{add:1} I would like to briefly address the more general questions
of why the quest to understand the mechanism responsible for superconductivity in the high $T_c$ cuprates is important and where are
we twenty years after Bednorz's and Muller's seminal discovery? As part of this, I will indicate how the numerical results for the
Hubbard model may help to provide some answers.

It has been suggested\cite{add:2} that the answer to the first question is that we need to ``domesticate the goat." That is, roughly
11,000 years ago man domesticated the wolf. During the next 4000 years, different wolf species as well as various breeds of dogs were
created and improved. But then, 7000 years ago the goat was discovered and domesticated. This achievement led within the next 1000 years to the
domestication of sheep, cattle, chickens, swine, etc.\ and to a fundamentally different way of life. So the high $T_c$ cuprate problem or, more generally, the problem of the strongly
correlated electron superconductors is the goat and we are seeking to domesticate it.  The hope is that this will lead to the
development of new strongly correlated superconducting materials and a deeper understanding of a variety of existing ones.

A short list of some of the materials\cite{add:3} discussed at the M2S-HTS Dresden meeting which fall into the category of strongly correlated superconductors follow:

\begin{description}
  \item [Cuprates] (hole and electron doped)
	\item [Heavy fermions] (U$_2$ (PdPt)$_3$ B, CeCoIn$_5$, PrOs$_4$ Sb$_{12}$, PuCoGa$_5$)
	\item [Ruthenates] Sr$_2$RuO$_4$
	\item [Organics] $\kappa$BEDT, (TMTSF)$_2$PF$_6$
	\item [Cobaltates] Na$_x$CoO$_2$ (1.3 H$_2$O)
\end{description}
%In addition, there are a number of other materials of interest because of their possible relationship to the cuprates such as the
%CuO ladder materials as well as interesting analogs such as Ce$_2$AgF$_4$.

In addressing the second question regarding where are we with respect to understanding the mechanism responsible for superconductivity
in the cuprates, it is useful to think back to a little over a decade ago. At that time a key question involved the symmetry of the gap.
At an APS March meeting, Bertrum Batlogg was asked when we would have an experiment which would tell us the symmetry of the cuprate gap.
His response was quick and to the point: We already have a number of such experiments, we just don't agree on which one is correct.
As it turned out, we were not so far away from settling this question. At that time, van Harlingen and his group at the University
of Illinois and Tsuei, Kirtley and their co-workers at IBM were carrying out phase sensitive experiments which would provide convincing
evidence that the cuprate gap was d-wave like (for orthorhombic systems d+$\alpha$s).

I believe that now, a decade later, the question regarding the nature of the cuprate pairing mechanism is at a similar stage.
That is, as seen from the talks presented at this conference, there are clearly a number of theoretical proposals for the underlying
cuprate pairing mechanism. Here is a short list, which I realize is incomplete (there were some 200 theoretical talks and 700 posters
covering electronic structure, many-body theories, phenomenology, and proposals for experiments and new materials):

\begin{description}
  \item Jahn-Teller bipolarons
	\item Stripes (the role of inhomogeneity)
	\item RVB-RMFT-Gutzwiller Projected BCS
	\item Electron-phonon+U
	\item Spin-fluctuations
	\item Charge-fluctuations
	\item Electric quadropole fluctuations
	\item Loop current fluctuations
	\item dDW, dCDW
	\item Quantum critical point fluctuations
	\item Competing phases
	\item Pomeranchuck instabilities
	\item d-to-d electronic excitations
\end{description}

The problem regarding the pairing mechanism is therefore reminiscent of the earlier gap symmetry question. It is not that there is a
lack of proposals, but rather it is that we do not have a consensus on which one contains the appropriate description.
Some of us had thought that the experimental observation of a d-wave gap, which had been predicted from the analysis of Hubbard and
t-J models, would have provided ``convincing" evidence that the pairing arose from a spin mediated interaction. But as we now know,
a $d_{x^2-y^2}$ gap is not a sufficiently unique signature. There are a number of possible pairing mechanisms such as B$_{1g}$ phonons,
fluctuating current loops, as well as others which may drive d-wave pairing.  In addition, there is the suggestion that the
pairing interaction is amplified by intrinsic inhomogeneities or stripes. There is also the question of the possible role of
amplification that may be obtained near a quantum critical point (QCP). Although, I believe that this latter QCP scenario can be viewed
as tuning the parameters to optimize a given pairing mechanism, much as what was done when $T_c$ was increased in the electron-phonon
systems by varying the composition in order to be near a lattice instability. In any event, the question remains, how will we know?
Let's consider some possibilities.

Perhaps it will be shown that there is another ordered phase associated with the pseudogap regime such as a d-DW or a d-CDW, a time
reversal breaking current-loop phase (recent neutron scattering and Kerr effect experiments), a non-superconducting xy-like phase
(non-analytic dependence of the magnetization as $\rm H\to O$ at temperatures above $T_c$), an exotic spin-liquid phase, \dots
There were in fact both theoretical and experimental talks on such phases, and definitive evidence for a new ordered phase would certainly narrow the
theoretical possibilities. Here however it is interesting to note that in spite of the evidence for a striped phase in La$_{1.35}$Sr$_{0.25}$Nd$_{0.4}$CuO$_4$
and increasing evidence for stripe-like fluctuations in the cuprates, the question of whether stripes, or other intrinsic inhomogeneities,
play an essential role in the high $T_c$ pairing mechanism remains open. We did hear about ARPES experiments on La$_{2-x}$Sr$_x$CuO$_4$ which found that the
gap peaks near $x=1/8$ where $T_c$ dips. Further experiments on dynamic inhomogeneity-induced pairing and the role of mesoscale structures
in the high $T_c$ problem are needed.

Perhaps we will know when we have a further understanding of the clues coming from the electronic structure calculations
regarding the variation of $T_c$ with the chemical structure and the effective band parameters such as the next near neighbor hopping
$t^\prime$. Perhaps the insight we need will come from an understanding of the local density of states measured by STM and its relation
to nearby local structural and chemical changes. Perhaps the ARPES, INS and STM studies will show that phonons play an important role.
If this is the case, it seems likely that it will not be in the traditional way they do in the low $T_c$ metals, but rather in combination with the stripes or mesoscale structures.

It may be, as suggested at this meeting, that one can identify a quantum critical point (QCP) which is associated with superconductivity
and then from the nature of the QCP, identify the fluctuations which mediate the pairing. Of course if the interplay between the various
phases is sufficiently strong, one may have to reconsider the meaning of mechanism.

One route that has been discussed is that of identifying the cuprate pairing mechanism by showing the similarity of the cuprates to
other classes of materials which exhibit unconventional superconductivity such as the actinide metals (PuCoGa$_5$, PuRhGa$_5$) and
the heavy fermion systems (CeRIn$_5$, R=Co,Rh,Ir). Perhaps there is a common mechanism which is
tunned as one moves from one system to another. The candidate mechanism discussed at this meeting was spin-fluctuations.

Another approach is based on using numerical techniques (or possibly experimental cold atom analogues) to study particular models
such as the Hubbard model. Here the idea is to determine first whether a given model exhibits the phenomena seen in the cuprates and
then, if it does, determine the nature of the pairing interaction in the model. (This is the approach taken in this chapter.)

Perhaps the best approach will be to follow the path taken for the traditional low temperature superconductors where the structure
of the pairing interaction was determined from the tunneling dI/dV characteristic. Here of course one knew how to get $\Delta(\omega)$
from dI/dV and had the Eliashberg theory to determine the interaction from $\Delta(\omega)$. For the cuprates one will likely need a
combination of numerical and analytic approaches along with ARPES, INS, STM and conductivity data to carry out such a program.

For the cuprates, the $k$ dependence of the gap near the fermi surface is known from a variety of experiments. A detailed map for
YBCO is now available from $\pi$-junction interference measurements which find an anisotropic d+$\alpha$s wave
form. If one had a simple ($\cos k_x-\cos k_y$) form over the entire Brillouin zone, one would know that the pairing arose from an
attractive near-neighbor interaction. While it is likely somewhat more extended and, for the case of an orthorhombic crystal, anisotropic,
the simple d-wave form is a reasonable starting point. What is needed are experimental determinations of the frequency dependence
of the gap. Actually one would like to determine both the renormalization parameter $Z(k,\omega)$ and $\Delta(k,\omega)=\phi(k,\omega)/Z(k,\omega)$
with $\phi(k,\omega)$ the gap parameter.  Studies aimed at extracting the frequency dependence of the gap and the underlying interaction using
conductivity $\sigma(\omega)$ as well as INS and ARPES data were reported. One can expect further progress in this direction as such
data becomes available on the same crystals.

%Ultimately, with so many years of work and so many materials, a theory of the cuprates will be judged by how well it unifies all of the
%observed phenomena. One is struck by the apparent similarity of the basic phenomena seen in these materials, and at the same time their
%remarkable sensitivity to structural and chemical details. Perhaps the very nature of the strong interaction leads to this, with materials
%which have different microscopic interactions having similar pseudopotentials when projected onto the low energy manifold. Then in this
%manifold, the competition between nearly degenerate states such as the striped state and the d-wave pairing state can be determined by
%relatively weak effects. This is what is seen in numerical studies of the Hubbard model.

So at the present time, the question of the mechanism responsible for pairing in the high $T_c$ cuprates remains open. However, it is
clear from the work presented at this conference that we are moving closer to an understanding.


\begin{thebibliography}{99}
\bibitem{ref:?} In the infinite cluster limit, one expects that $P_d$
will follow the low temperature Kosterlitz-Thouless behavior $P^{-1}_d\sim A\, \exp
(-2B/(T-T_c)^{0.5})$.
\bibitem{ref:Ref10} The eigenvalue for $U=4t$
on the 4-site cluster lays above the result obtained for the 24-site cluster.  As
discussed in Section~3, this is expected because the 4-site cluster encloses just one d-wave
plaquette and the embedding action does not contain pairfield fluctuations.
\end{thebibliography}
\begin{thebibliography}{99}
\bibitem{add:3} In addition, there were a number of talks about the sigma-orbital materials such as MgB$_2$, Li (high pressure form)
	and doped diamond, which are of great interest but are driven by electron-phonon interactions and do not appear to be directly related to the
	cuprate problem.
\end{thebibliography}
\end{document}